%% file: elsarticle-template-1-num.tex
\documentclass[preprint,12pt]{elsarticle}
\input{commands.tex}

\usepackage{amssymb}

\usepackage{lineno}

\journal{Renewable Energy}

\begin{document}

\begin{frontmatter}



\title{Fault Diagnosis of the 10MW Floating Offshore Wind Turbine Benchmark: a Mixed Model and Signal-based Approach}


\author[First]{Yichao Liu \corref{cor1}}
\author[First]{Riccardo Ferrari}
\author[Second]{Ping Wu}
\author[First]{Xiaoli Jiang}
\author[Third]{Sunwei Li}
\author[First]{Jan-Willem van Wingerden}

\address[First]{Delft University of Technology, Delft, 2628CD, The Netherlands}
\address[Second]{Zhejiang Sci-Tech University, Hangzhou, 310018, China}
\address[Third]{Graduate School at Shenzhen, Tsinghua University, Shenzhen, 518055, China}

\cortext[cor1]{Corresponding author. E-mail address: yichaoliu629@gmail.com}
\begin{abstract}
	\input{sections/0_abs}

\end{abstract}


\end{frontmatter}


\input{sections/Nomenclature}

\section{Introduction}
\input{sections/1_intro}
\label{sec:1}

\section{Description of the 10MW FOWT Benchmark} \label{sec:2}
\input{sections/2_description}

\section{Mixed FD Architecture for Floating Offshore Wind Turbines}\label{sec:3}
\input{sections/3_model_based}

\section{Case Study and Comparison}\label{sec:4}
\input{sections/4_sims}

\section{Conclusions}\label{sec:5}
\input{sections/5_final}

\section*{Acknowledgments}
This work was supported by the European Union via a Marie Sklodowska-Curie Action (Project EDOWE, grant 835901).


\bibliographystyle{model1-num-names}
\bibliography{references.bib}







\end{document}

%% file: commands.tex
\usepackage{graphicx,keyval,trig,url} 
\usepackage[T1]{fontenc}     
\usepackage[utf8]{inputenc}	
\usepackage{graphicx}

\usepackage{amsthm}
\usepackage{amssymb}
\usepackage{amsmath}
\usepackage{mathtools}
\usepackage[mathscr]{eucal}
\usepackage{lipsum}
\usepackage{cuted} 
\usepackage{blindtext}
\usepackage{color}
\usepackage{dsfont}

\usepackage{epstopdf}
\usepackage{bm}   

\usepackage{color}
\usepackage{xcolor}

\newtheorem{rem}{Remark}
\usepackage{multirow}
\usepackage{longtable}

\usepackage{nomencl}
\usepackage{etoolbox}

%% file: sections/0_abs.tex
Floating Offshore Wind Turbines (FOWTs) operate in the harsh marine environment with limited accessibility and maintainability. 
Not only failures are more likely to occur than in land-based turbines, but also corrective maintenance is more expensive. 
In the present study, a mixed model and signal-based Fault Diagnosis (FD) architecture is developed to detect and isolate critical faults in FOWTs.
More specifically, a model-based scheme is developed to detect and isolate the faults associated with the turbine system.
It is based on a fault detection and approximation estimator and fault isolation estimators, with time-varying adaptive thresholds to guarantee against false-alarms. 
In addition, a signal-based scheme is established, within the proposed architecture, for detecting and isolating two representative mooring lines faults.
For the purpose of verification, a 10MW FOWT benchmark is developed and its operating conditions, which contains predefined faults, are simulated by extending the high-fidelity simulator. 
Based on it, the effectiveness of the proposed architecture is illustrated. In addition, the advantages and limitations are discussed by comparing its fault detection to the results delivered by other approaches.
Results show that the proposed architecture has the best performance in detecting and isolating the critical faults in FOWTs under diverse operating conditions.

%% file: sections/Nomenclature.tex
\makenomenclature
\renewcommand\nomgroup[1]{%
  \item[\bfseries
  \ifstrequal{#1}{A}{\footnotesize Floating offshore wind turbine benchmark}{%
  \ifstrequal{#1}{B}{\footnotesize Model-based scheme}{%
  \ifstrequal{#1}{C}{\footnotesize Signal-based scheme}{%
  \ifstrequal{#1}{D}{\footnotesize Case study}{%
  }}}}%
]}
\nomenclature[A]{$k$}{Discrete time index}
\nomenclature[A]{$k_0$}{Unknown time index of faults occurrence}
\nomenclature[A]{$x$}{State vector}
\nomenclature[A]{$u$}{Controlled input vector}
\nomenclature[A]{$y$}{Measured output vector}
\nomenclature[A]{$A^0$}{Nominal linear part of the wind turbine healthy dynamics}
\nomenclature[A]{$\rho$}{Nominal nonlinear part of the wind turbine healthy dynamics}
\nomenclature[A]{$C^0$}{Nominal output matrix}
\nomenclature[A]{$\eta_x$}{Modelling uncertainties}
\nomenclature[A]{$\eta_y$}{Output disturbances}
\nomenclature[A]{$\beta$}{Time profile}
\nomenclature[A]{$C_l^0$}{$l$--th row of the nominal output matrix $C^0$}
\nomenclature[A]{$e_i$}{All-zeroes column vector of a suitable size having a single 1 in its $i$--th position}
\nomenclature[A]{$\phi_x$}{Process fault function}
\nomenclature[A]{$\phi_y$}{Output fault function}
\nomenclature[A]{$\vartheta_{y}$}{Parameter vector of the output fault function}
\nomenclature[A]{$\vartheta_{x}$}{Parameter vector of the process fault function}
\nomenclature[A]{$\Delta \rho$}{Change of the nonlinear part of the dynamics}
\nomenclature[A]{$\digamma$}{Fault class}
\nomenclature[A]{$\digamma_1$}{Fault class 1 being addressed by the model-based scheme}
\nomenclature[A]{$\digamma_2$}{Fault class 2 being addressed by the signal-based scheme}
\nomenclature[A]{$c$}{Scale factor of the rotor blade fault}

\nomenclature[B]{$A$, $B$, $C$, $D$}{System matrices of the fault detection and approximation estimator}
\nomenclature[B]{$L$}{Gain matrix of the fault detection and approximation estimator}
\nomenclature[B]{$\hat{x}$}{Predicted state vector}
\nomenclature[B]{$\hat{y}$}{Predicted output vector}
\nomenclature[B]{$\bar{\eta}_x$}{Bounding function of the process uncertainties}
\nomenclature[B]{$\bar{\eta}_y$}{Bounding function of the measurement uncertainties}
\nomenclature[B]{$\hat{\phi}_{x,0}$}{General online estimator}
\nomenclature[B]{$\hat{\Theta}^0$}{User designed parameters domain}
\nomenclature[B]{$M_{\hat{\Theta}^0}$}{Radius of the user designed parameters domain}
\nomenclature[B]{$\hat{\phi}_x$}{Linearly parameterized function for the actuator fault}
\nomenclature[B]{$\hat{\phi}_y$}{Linearly parameterized function for the sensor fault}
\nomenclature[B]{$\hat{\vartheta}_{x,l}$}{Parameter vector in the linearly parameterized function for the actuator fault}
\nomenclature[B]{$\hat{\vartheta}_{y,l}$}{Parameter vector in the linearly parameterized function for the sensor fault}
\nomenclature[B]{$g^l$}{Fault-specific shape function of the fault isolation estimator}
\nomenclature[B]{$\kappa^l$}{Maximum possible parameter estimation error of the threshold}
\nomenclature[B]{$l$}{Index of the estimators}
\nomenclature[B]{M}{Size of the output}
\nomenclature[B]{N}{Total number of the faults}
\nomenclature[B]{$r_{y,l}$}{Residuals of the $l$--th estimator}
\nomenclature[B]{$k_d$}{Fault detection time}
\nomenclature[B]{$\bar{r}_{y,l}$}{Threshold of the fault detection and approximation estimator}
\nomenclature[B]{$\alpha$, $\delta$}{Scalar constants of the threshold}
\nomenclature[B]{$\bullet_{(i)}$}{i--th component of a vector quantity}

\nomenclature[C]{$S_{\ddot{\alpha}}$}{Time-frequency spectrum of the tower-top acceleration}
\nomenclature[C]{$S_{\ddot{\alpha},q}$}{Training spectrum of the tower-top acceleration}
\nomenclature[C]{$\tau$}{Transformed time coordinate}
\nomenclature[C]{$w$}{Angular frequency coordinate}
\nomenclature[C]{$h(\bullet)$}{Moving window}
\nomenclature[C]{$Q$}{Total number of the training spectra}
\nomenclature[C]{$P$}{Total number of the frequency bins}
\nomenclature[C]{$w_p$}{Frequency bin}
\nomenclature[C]{$K$}{Neighborhood size}
\nomenclature[C]{$D_{K,q}$}{Distance between the training spectra and the actual one}

\nomenclature[D]{$U_m$}{Constant wind speed}
\nomenclature[D]{$U_f$}{Fluctuating wind component}
\nomenclature[D]{$H_s$}{Significant wave height}
\nomenclature[D]{$T_p$}{Peak-spectral period}
\nomenclature[D]{$T_s$}{Discrete time step}
\nomenclature[D]{$\theta_{\mbox{in}}$}{Control input of the pitch angle}
\nomenclature[D]{$T_{g, \mbox{in}}$}{Control input of the generator torque}

\printnomenclature
\color{black}

%% file: sections/1_intro.tex
The penetration of wind power into the energy mix has been significantly growing over the past decade {\cite{Decastro-2019, Francisco_2020}}. In 2018, the global installed capacity of wind energy reached 591,549MW worldwide \cite{GWEC-2019}.
Following in the path of onshore wind exploitation, the developmentment of offshore wind energy picks up the momentum in the race in transitioning from conventional fossil fuels to renewable energy. In particular, offshore wind turbines are less intrusive from a visual and acoustic point of view, and guarantee much higher and steady generation of power \cite{Liu-2016}.
As the exploitation of offshore wind energy moves from shallow to deep waters, Floating Offshore Wind Turbines (FOWTs) become the ideal alternative to replace the bottom-fixed turbines and capture the abundant wind resources available over the deep sea \cite{Liu-2016}. Although FOWTs are evidently advantageous from the public acceptance and power generation point of view, designing, operating and maintaining them pose a series of challenges:

\begin{enumerate}
\item FOWTs may experience unexpected faults and failures due to the complex environmental loads, which would in turn lead to operation interruption, economic losses and thus high Operation and Maintenance (O\&M) costs \cite{Cho-2018}. 
\item The reliability of FOWTs decreases with increasing turbine size and complexity \cite{Faulstich-2011}. In particular, FOWTs tend to have larger rotor diameters than land-based counterpart, and hence structural faults of the rotor blade are more prominent in FOWTs.
\item FOWTs are usually situated in deep waters at a considerable distance from the shore, which restrict their accessibility and maintainability \cite{Levitt-2011}. As a consequence, the O\&M cost of a FOWT may account for as high as 30\% of the total life cycle cost \cite{Dinwoodie-2013}, and also is an influential factor in the determination of the turbine's cost of energy. 
\end{enumerate}

To assess the reliability of wind turbines and to reduce the O\&M cost, several Fault Diagnosis (FD) architectures have been proposed in the literature \cite{Saha-2019}.
In general, FD architectures employ sensor data for detection, isolation and identification of the faults in the wind turbine components, whose output can be used to implement fault tolerant control and condition-based maintenance \cite{Ding-2008}.
Odgaard \textit{et al}. \cite{Odgaard-2009} introduced a 4.8MW bottom-fixed wind turbine benchmark in 2009. Using this benchmark, different FD approaches were implemented and verified, including a Kalman filter combined with a diagnostic observer approach \cite{Chen-2011}, up-down counter \cite{Ozdemir-2011}, Gaussian-kernel support vector machine \cite{Laouti-2011} and estimation-based method \cite{Zhang-2011}. These approaches were summarized comprehensively by Odgaard \textit{et al}. \cite{Odgaard-2013}. In fact, it has been found by Odgaard \textit{et al}. \cite{Odgaard-2009} that most faults in the benchmark can be detected successfully by the above mentioned approaches.
Recently, several attempts have been made to build FD architectures via alternative approaches, such as interval observer \cite{Sanchez-2015}, sliding mode observer \cite{Hosseinzadeh-2016} and multi-physics graphical model \cite{Mojallal-2018}.

Nevertheless, the existing FD architectures which have been developed for land-based wind turbines can not be directly applied to the FOWT, the reasons for this include:
\begin{enumerate}
\item There is no systematic study, to the best of authors' knowledge, addressing prominent faults of FOWTs, such as structural faults of the rotor blades and of the mooring lines. More specifically, the published works mainly focus on the pitch system, gearbox, bearings, and electrical system \cite{Odgaard-2009,Odgaard-2013} only. 
One plausible explanation might be that there is currently no benchmark including the fault scenarios of the rotor blades and mooring lines, nor those are implementable in widely-used simulation packages such as Fatigue, Aerodynamics, Structures, and Turbulence (FAST) \cite{Jonkman-2005}. 

\item The existing FD architectures are, in general, based on a simplified model of the land-based wind turbine \cite{Odgaard-2009,Odgaard-2013}, which lacks the descriptions of hydro and mooring line dynamics. 
However, the dynamics of the FOWT system are significantly influenced by the complex interactions among the floating foundation, the mooring system, and the rotor blades \cite{Chen-2019}.  
\end{enumerate}

Therefore, it is urgent to develop an effective FD architecture for FOWTs, considering the inapplicability of existing architecture, increased system size and complexity of turbine-foundation interaction and reduced accessibility for maintenance.

Recently, some preliminary efforts have been made to develop FD architectures for selected FOWT components. For instance, Ghane \textit{et al}. \cite{Ghane-2018} utilized the statistical approach to detect and estimate the wear in the drivetrain of the 5MW spar-type FOWT. 
Cho \textit{et al}. \cite{Cho-2018} used the model-based approach to detect, isolate and accommodate the faults of the blade pitch system of the same FOWT. 
Zhang \textit{et al}. \cite{zhang-2018} employed the data-driven approach, namely random forests, to detect several actuator and sensor faults of the 5MW semi-submersible FOWT. 
In spite of these designated contributions, relatively few studies addressed aforementioned challenges.

Considering the lack of systematic studies concerning the FD architecture for FOWTs, the present study aims to develop an effective approach to detect and isolate the fault of the FOWTs, and therefore contributes in the following aspects,

\begin{enumerate}
\item Instead of considering one single component, e.g. blade pitch system \cite{Cho-2018}, of the FOWT system, a mixed FD architecture is developed in the present study for the detection and isolation of faults in both the turbine and mooring systems in FOWTs. 
In particular, the proposed mixed FD architecture is established by integrating the model-based scheme suggested by Ferrari \textit{et al}. \cite{Ferrari-2008} with a signal-based scheme.

\item In order to verify the proposed architecture, a novel 10MW FOWT benchmark is set up based on the DTU three-bladed variable speed reference wind turbine and Triple-Spar floating foundation \cite{fontanella2018linear,fontanella2018control}. 
Several critical fault scenarios including not only actuator and sensor faults, but also structural faults of rotor blade and mooring lines, are generated for the FD purpose.
Particularly, the models of actuator and sensor faults are extracted and improved from the models available in the literature \cite{Odgaard-2013, Badihi-2015}.
\end{enumerate}

{In summary, the main contribution of this study is the FOWT-oriented mixed FD architecture, which integrates model and signal-based schemes to encompass most critical faults in FOWTs.
Another novelty is the physics-based FOWT benchmark. 
For the first time, a FOWT benchmark is established for the purpose of developing and verifying a FOWT-oriented FD architecture. In particular, the structural faults of the rotor blades and of the mooring lines are taken into consideration.}

From detecting and isolating the predefined faults in the developed 10MW FOWT benchmark, the effectiveness of the proposed FD architecture is illustrated. In addition, two classic fault detection methods are applied to the 10MW FOWT benchmark with predefined faults. The comparison concerning the fault detection hence reveals the advantages and limitations of the proposed mixed FD architecture. 

The remainder of the paper is organized as follows. Section \ref{sec:2} introduces the 10MW FOWT benchmark. In section \ref{sec:3}, the overall structure and theoretical framework of the mixed FD architecture are presented. A case study employing the proposed method to discern the faults of the newly developed 10MW FOWT benchmark is given in section \ref{sec:4}. Then, its advantages and limitations are discussed by comparing the detection results to other classical approaches. Section \ref{sec:5} contains concluding remarks.



%% file: sections/2_description.tex
\begin{figure}
\centering \includegraphics[width=0.8\columnwidth]{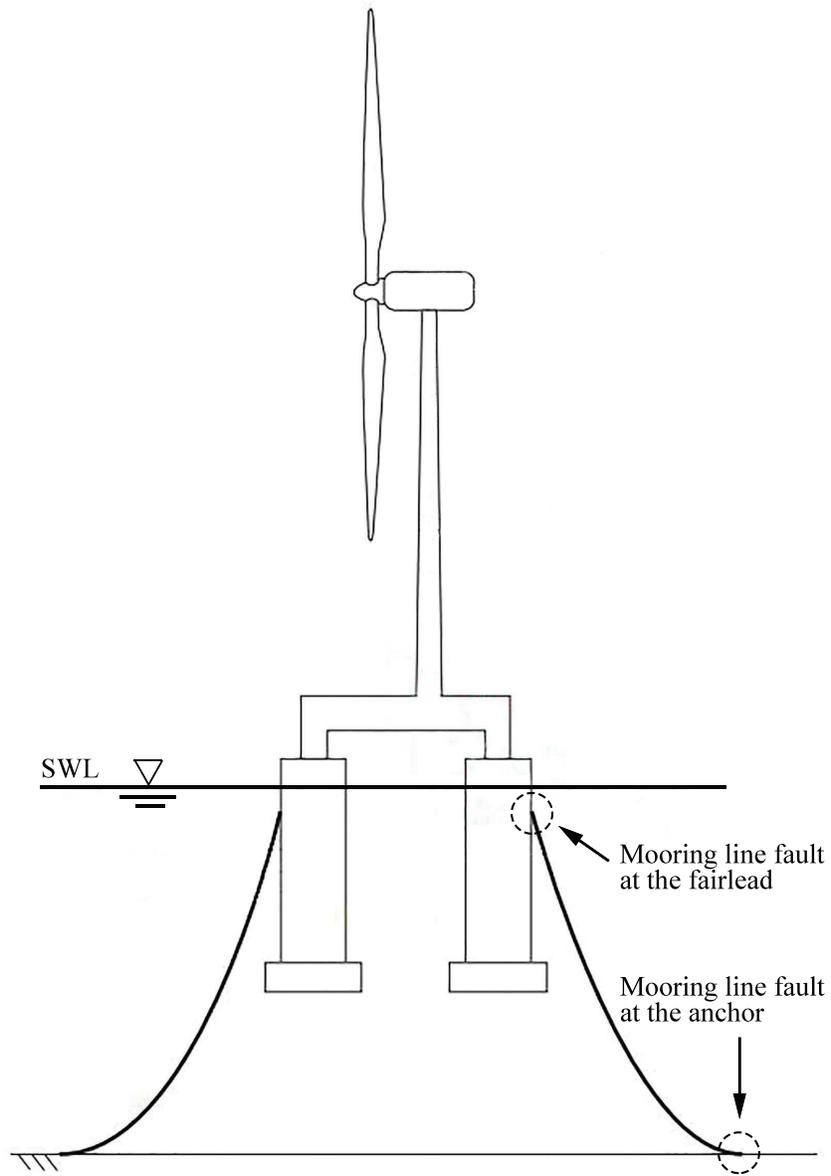}
\caption{Lateral view of the DTU 10MW wind turbine, of the TripleSpar floating platform and indication of mooring line faults location.}
\label{Pic_drawing} %
\end{figure}

In this section, the 10MW FOWT benchmark is presented. It is based on the DTU three-bladed variable speed reference wind turbine and the Triple-Spar floating platform \cite{fontanella2018linear,fontanella2018control}.
Particularly, several critical fault scenarios, including not only the actuator and sensor faults, but also the structural faults, are defined and implemented in the benchmark. 
The overview of the FOWT is portrayed in Fig.~\ref{Pic_drawing} while its specifications are listed in Table~\ref{table:FOWT}. 

\begin{table}
\setlength{\tabcolsep}{0.05mm}{
\caption{Specifications of the 10MW FOWT {(SWL: Sea Water Level)}.\label{table:FOWT}}
\begin{tabular}{ll}
\hline 
Parameter     & Value            \\ \hline
\textbf{Turbine system}           \\
Rating           & {10\, MW}          \\
Rotor orientation, configuration  & Upwind, 3 blades            \\
Pitch control         & Variable speed, collective pitch        \\
Drivetrain      & Medium speed, multiple stage gearbox          \\
Rotor, hub diameter & {178.3\,m, 5.6\,m} \\
Hub height & {119\,m} \\
Cut-in, rated, cut-out wind speed & {4\,m/s, 11.4\,m/s, 25\,m/s} \\
Cut-in, rated rotor speed & {6\,rpm, 9.6\,rpm} \\
Rated tip speed & {90\,m/s} \\ \hline
\textbf{Floating platform}  \\
Total height and draft & {66\,m, 56\,m} \\
Distance from the tower center-line & {26\,m} \\
Single column diameter & {15\,m} \\
Column elevation above SWL & {10\,m} \\
Elevation of tower base above SWL & {25\,m} \\
Water displacement & {29497.7\,m$^3$} \\ \hline
\textbf{Mooring lines} \\
Number of lines & 3 \\
Line angles from upwind direction & {0\,$^{\circ}$, 120\,$^{\circ}$, 240\,$^{\circ}$} \\
{Water depth and anchor radius} & {180\,m, 599.98\,m} \\
Fairleads above SWL & {8.7\,m} \\
Fairleads radius & {47.181\,m} \\
Line diameter & {0.18\,m} \\
Total length & {707\,m} \\
Mass/length in air & {594\,kg/m} \\ \hline 
\end{tabular}}
\end{table}

For the purpose of developing the proposed mixed FD architecture, {the dynamics of the 10MW FOWT and of its actuators in state-space will be described}, by means of the following discrete-time system
\begin{align}
\begin{cases}
 x(k+1)  \! \! \! &= A^0x(k)+\rho(x(k),u(k))+\beta(k-k_0)\times \\
& \;\;\; \phi_x(y(k),u(k),\vartheta_x)+\eta_x(x(k),u(k),k)  \\
 y(k) \! \! \! &= C^0x(k)+\beta(k-k_0)\phi_y(x(k),\vartheta_y)+\\
& \;\;\; \eta_y(x(k),u(k),k)
\end{cases} \, ,
\label{eq:FOWT_DYNAMICS}
\end{align}
where $k=0,1,\dots$ is the discrete time index and $x\in\mathbb{R}^n$, $u\in\mathbb{R}^m$, $y\in\mathbb{R}^p$ represent the state, the controlled input and the measured output vectors, respectively. 
The contents of $u$ and $y$ are defined in Table~\ref{table:FOWT_vars}. The matrix $A^0\in\mathbb{R}^{n\times{n}}$ and the vector field $\rho:\mathbb{R}^n\times\mathbb{R}^m\mapsto\mathbb{R}^n$ denote the nominal linear and nonlinear parts of the FOWT healthy dynamics while $C^0\in\mathbb{R}^{p\times{n}}$ is the nominal output matrix. 
The unavoidable modelling uncertainties and output disturbances are described by the functions $\eta_x:\mathbb{R}^n\times\mathbb{R}^m\times\mathbb{R}\mapsto\mathbb{R}^n$ and $\eta_y:\mathbb{R}^n\times\mathbb{R}^m\times\mathbb{R}\mapsto\mathbb{R}^p$.

The terms $\beta(k-k_0)\times\phi_x(y(k),u(k),\vartheta_x)$ and $\beta(k-k_0)\times \phi_y(x(k),\vartheta_y)$ represent the dynamic changes of the state and output equation, respectively, due to the occurrence of faults at the unknown time index $k_0$, similarly to the case reported by Ferrari \textit{et al}. \cite{Ferrari-2008}.
\begin{table}
\caption{Controlled and measured variables of the 10MW FOWT.\label{table:FOWT_vars}}
\begin{center}
\begin{tabular}{lcl}
\hline 
Variable     & Symbol   & Units           \\ \hline
\textbf{Control signals}           \\
Reference pitch angle           & $\theta_r \equiv u_1$ & deg   \\
Reference generator torque & $ T_{g,r} \equiv u_2  $  &  Nm  \\\hline
\textbf{Measurements}  \\
Pitch angles & $\theta_s \equiv y_s,\, s=1, \dots, 3$ & deg   \\
Generator torque & $T_g \equiv y_4$ & Nm   \\
Rotor angular speed & $\omega_r \equiv y_5$ & rad/s   \\
Generator angular speed & $\omega_g \equiv y_6$ & rad/s  \\
Generator power & $P_g \equiv y_7$ & W      \\
Blade root bending moments\protect\footnotemark & $M_{cm,s} \equiv y_{7+s},\, s=1, \dots, 3$ & Nm  \\
Tower top acceleration & ${\ddot{\alpha}} \equiv y_{10}$ & m/s$^2$ \\\hline 
\end{tabular}
\end{center}
\end{table}
In detail, a fault class $\digamma = \{\digamma_1, \, \digamma_2 \}$ consisting of 5 process and 6 output fault functions will be considered, with $\digamma_1$ being addressed by the model-based scheme, and $\digamma_2$ by the signal-based scheme in the proposed mixed FD architecture:
\begin{equation}
\digamma \triangleq \{\underbrace{\phi_{y,1}, \, \dots, \,\phi_{y,4},\phi_{y,7},\phi_{y,8} \,
\phi_{x,5}, \, \phi_{x,6} , \,\phi_{x,9},\,}_{\digamma_1}
\underbrace{\phi_{x,10}, \,\phi_{x,11}}_{\digamma_2}\} \, .
\label{eq:fault_set}
\end{equation}
The details of each fault function in $\digamma$ are described in Table~\ref{table:fault_function}.

\begin{table}
\setlength{\tabcolsep}{1mm}{
\caption{Fault types and functions.\label{table:fault_function}}
\begin{tabular}{llcc}
\hline 
Name  & Type & Description  & Function \\ \hline
\textbf{Sensor}\\
\vspace{2pt}
$f_1$  & scaling & Generator speed & $\phi_{y,1}=\vartheta_{y,1}C_1^0x$  \\
\vspace{2pt}
$f_2$  & scaling & Generator power & $\phi_{y,2}=\vartheta_{y,2}C_2^0x$  \\
\vspace{2pt}
$f_3$  & offset & Blade \hspace{-4pt} root \hspace{-4pt} bending \hspace{-4pt} moment & $\phi_{y,3}=\vartheta_{y,3}e_3$ \\
\vspace{2pt}
$f_4$  & scaling & Rotor speed & $\phi_{y,4}=\vartheta_{y,4}C_4^0x$ \\
\vspace{2pt}
$f_7$  & stuck & Pitch sensor & $\phi_{y,7} \! = \! \vartheta_{y,7} \!- \! C_5^0x \! - \! \eta_{y,(5)}$ \\
\vspace{2pt}
$f_8$  & scaling & Torque sensor & $\phi_{y,8}=\vartheta_{y,8}C_6^0x$  \\\hline
\textbf{Actuator} \\
\vspace{2pt}
$f_5$  & stuck & Pitch actuator & $\phi_{x,5}=-u+\vartheta_{x,5}e_5$  \\
$f_6$  & offset & Torque actuator & $\phi_{x,6}=\vartheta_{x,6}e_6$  \\\hline
\textbf{Structural} \\
\vspace{2pt}
$f_9$  & {parameter} & Rotor blade sudden fault  & $\phi_{x,9}= \Delta \rho(x,u,\vartheta_{x,9})$ \\
\vspace{2pt}
$f_{10}$ & {parameter} & Mooring line (fairlead) & $\phi_{x,10}= \Delta \rho(x,u,\vartheta_{x,10})$ \\
$f_{11}$ & {parameter} & Mooring line (anchor)  & $\phi_{x,11}= \Delta \rho(x,u,\vartheta_{x,11})$      \\ \hline 
\end{tabular}}
\end{table}

\footnotetext{$M_{cm,s}$ is the transformed symmetric moment of the $s$--th ($s=1, \, 2, \,3$) blade root via the Coleman transformation \cite{Wei-2008}, which transforms from a rotational coordinate frame to a fixed one where all periodic parts vanish.}

The parameters $\vartheta_{y,l}, \, l\in{1, 2, 4, 8}$ denote the scaling error magnitude for sensors affected by the scaling type of faults, while $\vartheta_{y,7}$ and $\vartheta_{y,3}$ denote, respectively, the stuck and offset value for these kinds of sensor faults. Similarly, $\vartheta_{x,5}$ and $\vartheta_{x,6}$ denote the same for actuator type of faults. 

The last three faults, described by $\phi_{x,9}$ to $\phi_{x,11}$, denote structural faults that affect the nonlinear dynamic function $\rho$. In order to describe such structural faults, the notations $\Delta \rho(x,u,\vartheta_{x,l}) \triangleq \rho(x,u,\vartheta_{x,l}) - \rho(x,u)$, with $l=\,9,\,\dots,\,11$, are introduced. They represent the change in the nonlinear part of the dynamics from its nominal behaviour to a faulty behaviour characterized by the parameter vectors of $\vartheta_{x,l}$. These vectors contain specific values of structural parameters which are caused by the structural faults. 

In particular, the faulty blade stiffness $\vartheta_{x,9} = c\cdot \vartheta_{x,9}^0$ is used to describe the abrupt rotor blade faults due to the effects of cracking, debonding/delamination or fiber breakage \cite{Pawar-2007}. 
In equations shown in Table~\ref{table:fault_function}, 
$0 \leq c < 1$ is a scale factor quantifying the reduction of the blade stiffness.
{
Usually, the stiffness reduction of the blade fault can be divided into three stages \cite{Degrieck_2001}. The first two stages are attributed to non-severe blade faults with a local stiffness reduction of around $0\%-30\%$. They are caused by transverse matrix cracks and debonding/delamination, respectively, which exert little effect on the global bending stiffness.
The third stage is severe blade fault with a local stiffness reduction of at least $30\%$.
It is mostly induced by the local damage progression and the fiber breaking, and may have a significant effect on the global dynamics of the blade.
In this paper, only the severe blade fault in stage 3 is taken into account for investigations.}

In order to properly simulate the abrupt rotor blade faults, the corresponding faulty mode shapes of the blades, which are dependent on $\vartheta_{x,9}$, are calculated using the tool Modes\footnote{Modes: a simple mode-shape generator for towers and rotating blades. https://nwtc.nrel.gov/Modes.} and fed into the FAST simulator. 
Note that the last two functions describe two critical mooring line faults (see Fig. \ref{Pic_drawing}) that may occur during the FOWT operation process \cite{Thomas-2018, Liu-2019}. 
In detail, the top segment fault at the fairlead will cause the mooring line to fall away and reduce its tension $\vartheta_{x,10}$ to zero. 
In the case of bottom segment faults, the static friction forces from the seabed fail to keep the anchor in balance. As a result, the anchor moves into a new equilibrium position where the unstretched length $\vartheta_{x,11}$ of the mooring line increases.

Based on the benchmark described above, the 10MW FOWT dynamics are simulated by the FAST simulator with Simulink.
In particular, all the fault functions listed in Table \ref{table:fault_function}, describing all the fault scenarios in $\digamma$, are included into the 10MW FOWT benchmark by extending the source code of the FAST simulator.
More importantly, critical parameters of the fault functions, e.g. magnitude of the faults and changed mode shapes of the blades, can be determined by users through an interface in Simulink, and thereby fed into the FAST simulator for the fault generation. 
As presented in Fig. \ref{Pic_block}, the lower part of the block diagram illustrates the developed 10MW FOWT benchmark. 
In addition, the well-known FAST numerical package \cite{Jonkman-2005} is customized to include all kinds of faults and then simulate the dynamics of a FOWT system containing specific faults. 
Furthermore, the control laws which are based on Linear Time Invariant (LTI) dynamical systems, are implemented in Simulink \cite{fontanella2018linear,fontanella2018control}.

%% file: sections/3_model_based.tex

\begin{figure}
\centering \includegraphics[width=1\columnwidth]{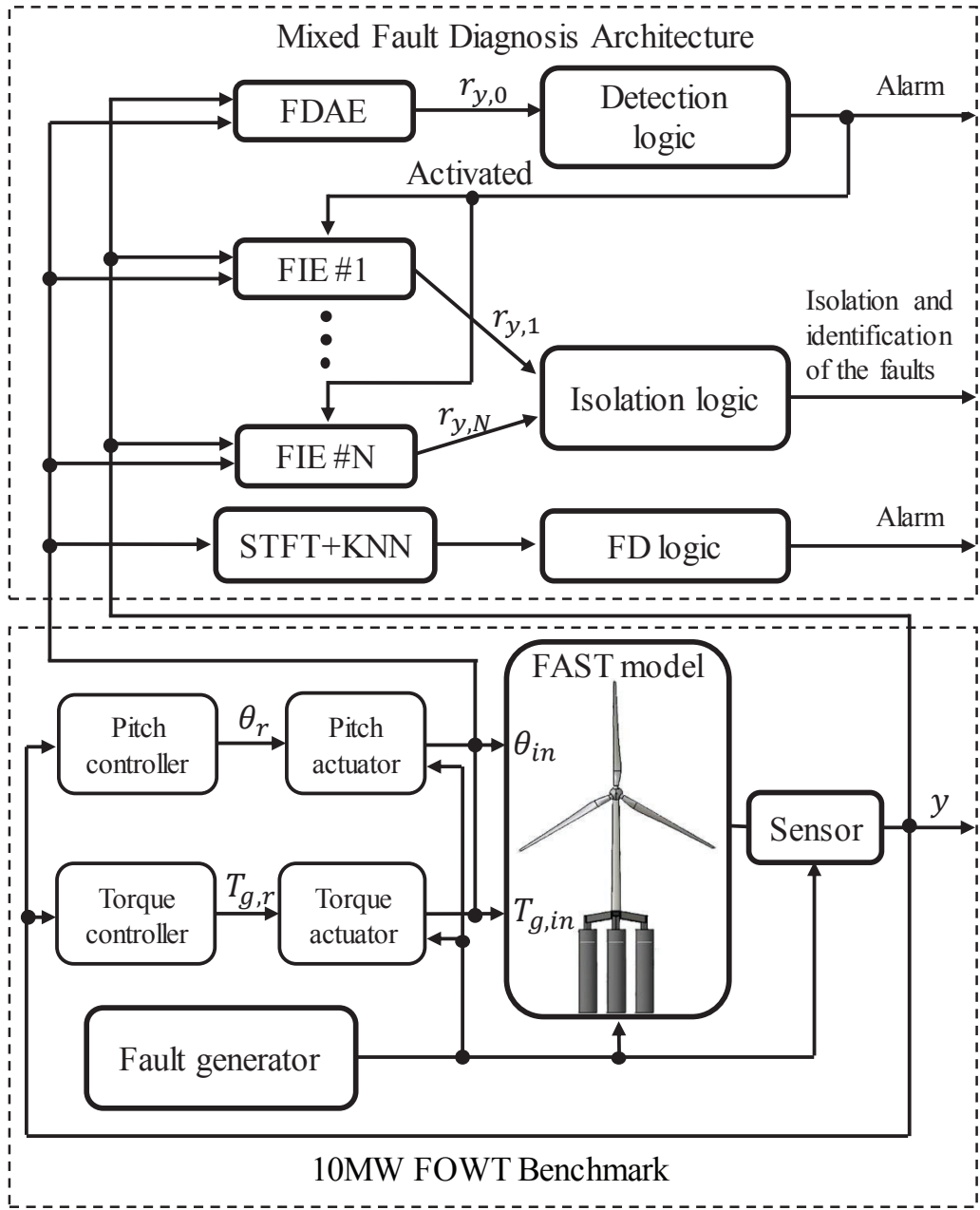}
\caption{Block diagram of the 10MW FOWT benchmark in the lower part and the overall structure of the mixed FD architecture (eqs.\eqref{eq:observer}-\eqref{eq:KNN}) in the upper part. A fault generator block is added to allow introducing faults (Table~\ref{table:fault_function}) in actuators, sensors, rotor blades and mooring lines.}
\label{Pic_block} %
\end{figure}

In this section {the overall structure of the mixed FD architecture is introduced}, which is depicted in the upper part of Fig. \ref{Pic_block}.
The mixed FD architecture includes a model-based FD approach in the time domain, which follows the generalized observer scheme \cite{patton1989fault}. 
In detail, a Fault Detection and Approximation Estimator (FDAE) is used to detect faults and activate a bank of $N$ Fault Isolation Estimators (FIEs), with $N$ being the number of the predefined faults. Fault decisions produced by the FDAE and the FIEs are evaluated by a detection and isolation logic in order to produce a final diagnosis.

For the purpose of coping with the faults associated with mooring lines, which proved to be elusive for estimator-based detectors, the proposed mixed FD architecture {includes} as well a signal-based FD scheme in the frequency domain. The signal-based scheme consists of a Short Time Fourier Transform (STFT) step followed by a K--Nearest Neighbour (KNN) detection and isolation step.

\subsection{Model-based fault diagnosis for the turbine system}
To detect and isolate the $N=9$ faults belonging to $\digamma_1$, a bank of $N+1$ nonlinear adaptive estimators is designed and implemented, each one yielding an output estimate $\hat{y}_{l}\in\mathbb{R}^M, \, l=0,...,N$, where $M$ is the size of the output. The FDAE ($l=0$) detects known faults and approximate unknown ones, while the FIEs ($l=1, \dots, N$), each corresponding to one fault in $\digamma_1$, are designed to isolate the detected fault. For each $l$--th estimator a residual signal is defined as
\begin{equation}
\begin{gathered}
r_{y,l}(k)=y_{l}(k)-\hat{y}_{l}(k) \, .
\end{gathered}
\label{eq:r}
\end{equation}
The healthy hypothesis ($l=0$) shall be rejected by the FDAE if the absolute value of at least one component $r_{y,l,(i)}(k)$ of its residuals, with $i\in \{1,\,\dots,\,,M\}$, exceeds the corresponding one of a suitable time-varying threshold, denoted as $\bar{r}_{y,l,(i)}(k)$. The first such time will be indicated as the \emph{detection instant} $k_d$. Similarly, each FIE ($l=1, \dots, N$) will reject its own faulty hypothesis if a component of its residuals will cross its threshold. If at some time instant every hypothesis but the $l$--th one has been rejected by the FDAE and FIEs, then the detection and isolation logic will conclude that the $l$--th fault occurs.

\noindent \subsubsection{FDAE and fault detection}
The FDAE will be based on a linearized version of the nominal healthy FOWT dynamics introduced in eq.~\eqref{eq:FOWT_DYNAMICS}.
For time instants $0\le k< k_d$, that is before detection, it will be described by the following LTI system 
\begin{equation}
\begin{cases}
\hat{x}(k+1) \!\! &= A\hat{x}(k)+Bu(k)+L(y(k)-\hat{y}(k)) \\
\hat{y}(k) \!\! &= C\hat{x}(k)+Du(k)
\end{cases} \, ,
\label{eq:observer}
\end{equation}
where $\hat{x}(k)$ and $\hat{y}(k)$ are the predicted state vector and output, respectively, and the FDAE input and output correspond to the same variables described in Table~\ref{table:FOWT_vars}. Following the block diagram introduced in the lower part of Figure~\ref{Pic_block}, the matrices $A$, $B$, $C$ and $D$ are obtained by cascading the LTI systems describing the actuator dynamics \cite{fontanella2018linear,fontanella2018control} with a linear model of the turbine. Such a linear model is obtained by applying subspace identification \cite{Gestel-2001} techniques to simulation data produced by FAST. It should be noted that it is rather difficult to obtain a white-box model due to the nonlinear dynamics of the FOWT \cite{Bakdi-2019}, as detailed in Section \ref{sec:4}. 
The matrix $L\in\mathbb{R}^{K\times{M}}$ is the FDAE gain, which is chosen such that $A_0\triangleq{A-LC}$ is stable.
The time-varying threshold $\bar{r}_{y,0}(k)$ is then designed to bound the healthy residual $r_{y,0}(k)$, in order to guarantee no false-positive alarms. 

In detail, the $i$--th component $\bar{r}_{y,0,(i)}(k)$ is calculated as,
\begin{equation}
\begin{gathered}
\bar{r}_{y,0,(i)}(k)\triangleq \sum_{h=0}^{k-1}\alpha_{(i)}\delta_{(i)}^{k-1-h}[\bar{\Delta} { \rho}(h)+\bar{\eta}_x(h)]+\alpha_{(i)}\delta^k_{(i)}\bar{\epsilon}^0_x(0)\\
+\bar{\eta}_{y,(i)}(k), i=1,...,M \, ,
\end{gathered}
\label{eq:14}
\end{equation}
where two scalar constants $\alpha_{(i)}$ and $\delta_{(i)}$ are obtained as in \cite{Zhang-2001}. $\bar{\eta}_x$ and $\bar{\eta}_y$ are the bounding function of the uncertainties in the state and output equations of the FOWT. The function $\Delta{ \rho}$ represents the difference between the nominal nonlinear healthy dynamics and the linear model:
\begin{equation}
\begin{cases}
\Delta{ \rho}(x(k),\hat{x},u(k))\triangleq { \rho}(x(k),u(k))-{ \rho}(\hat{x}(k),u(k)) \\
\bar{\Delta}{ \rho}(\hat{x}(k),u(k)) \;\;\; \triangleq \max\limits_{x\in{R^x}} (\|\Delta{ \rho}(x,\hat{x},u(k))\|)
\end{cases} \, .
\label{eq:difference of dynamics}
\end{equation}


If the condition  $|r_{y,0,(i)}(k)|>\bar{r}_{y,0,(i)}(k)$ holds for some component $i$ at the time instant $k=k_d$, a fault is detected and the FDAE equation becomes 
\begin{equation}
\begin{cases}
\hat{x}(k+1) &= A\hat{x}(k)+Bu(k)+L(y(k)-\hat{y}(k))+\\
& \; \; \; \hat{\phi}_{x,0}(y(k),u(k),\hat{\vartheta}_0(k)) \\
\hat{y}(k) &= C\hat{x}(k)+Du(k) \, ,
\end{cases}
\label{eq:approximator}
\end{equation}
where $\hat{\phi}_{x,0}$ is a general online estimator to learn unknown or unstructured faults (such as the rotor blade sudden fault $f_9$). For instance, $\hat{\phi}_{x,0}$ can be based on a Radial Basis Function (RBF) neural network, whose parameters are assumed to satisfy $\hat{\vartheta_0}\in\hat{\Theta}^0$, where $\hat{\Theta}^0$ is a user designed parameters domain. In particular, $\hat{\Theta}^0$ is chosen here to be an origin-centered hyper-sphere with radius of $M_{\hat{\Theta}^0}$. 

\subsubsection{FIE and fault isolation}
After a fault is detected at $k$=$k_d$, the bank of FIEs is activated. The actuator fault $l$--th FIE's dynamics, for the actuator faults $l=5,6$ are,
\begin{equation}
\begin{cases}
\hat{x}^l(k+1) \! \! \! &= A\hat{x}^l(k)+Bu(k)+L(y(k)-\hat{y}^l(k))+\\
& \; \; \; \hat{\phi}_{x,l}(y(k),u(k),\hat{\vartheta}_{x,l}(k)) \\
\hat{y}^l(k) \! \! \! &= C\hat{x}^l(k)+Du(k)
\end{cases} \, ,
\label{eq:isolation_x}
\end{equation}
where $\hat{\phi}_{x,l}(k)\triangleq col(\hat{\phi}_{x,l,(i)}(k), i=1,...,M)$ and $\hat{\vartheta}_{x,l}(k) \triangleq col(\hat{\vartheta}_{x,l,(i)}(k))$.
Particularly, 
$\hat{\phi}_{x,l,(i)}(y(k),u(k),\hat{\vartheta}_{x,l,(i)}(k)) \triangleq \hat{\vartheta}_{x,l,(i)}(k)^T g^l_{(i)}(y(k),u(k))$ is a linearly parameterized function. $\hat{\vartheta}_{x,l,(i)}(k)\in\hat{\Theta}^l_{(i)}$ denotes the parameter vector and $g^l_{(i)}$ is the fault-specific shape function. $\hat{\Theta}^l_{(i)}$ is assumed to be an origin-centered hyper-sphere with radius of $M_{\hat{\Theta}^l_{(i)}}$ as well.

In order to learn the fault function of $\phi_{x,l}$, the parameters of $\hat{\phi}_{x,l}$ are updated following the learning law,
\begin{equation}
\begin{gathered}
\hat{\vartheta}_{x,l}(k+1)=P_{\hat{\Theta}^l}(\hat{\vartheta}_{x,l}(k)+\gamma^{l}(k)g^l(k)C^T r^l_y(k+1)) \, ,
\end{gathered}
\label{eq:9}
\end{equation}
where $P_{\hat{\Theta}^l}$ denotes the projection operator on $\hat{\Theta}^l$ as
\begin{equation}
\begin{gathered}
P_{\hat{\Theta}^l}(\hat{\vartheta}_{x,l})\triangleq 
\begin{cases}
\hat{\vartheta}_{x,l}\, ,  \;
 \mbox{if} \; \;\;|\hat{\vartheta}_{x,l}|\leq M_{\hat{\Theta}^l} \\
(M_{\hat{\Theta}^l}/|\hat{\vartheta}_{x,l}|)\cdot \hat{\vartheta}_{x,l} \, , \; 
\mbox{if} \;\;\; |\hat{\vartheta}_{x,l}|> M_{\hat{\Theta}^l}
\end{cases}
\, .
\end{gathered}
\label{eq:projection}
\end{equation}
In addition, $\gamma^l_{(i)}(k)$ is calculated as,
\begin{equation}
\begin{gathered}\gamma^l(k)\triangleq \mu^l/(\varepsilon^l+\|g^l(k)C^T\|^2_F), \varepsilon^l>0, \quad  0<\mu^l<2 \, .
\end{gathered}
\label{eq:gamma}
\end{equation}

Similarly, the dynamics of the $l$--th FIE corresponding to the sensor faults, $l\in \{1,2,3,4,7,8\}$, are
\begin{equation}
\begin{cases}
\hat{x}^l(k+1) &= A\hat{x}^l(k)+Bu(k)+L(y(k)-\hat{y}^l(k))\\
\hat{y}^l(k) &= C\hat{x}^l(k)+Du(k)+\\
& \; \; \; \hat{\phi}_{y,l}(x(k),\hat{\vartheta}_{y,l}(k))
\end{cases} \, ,
\label{eq:isolation_y}
\end{equation}
where $\hat{\phi}_{y,l}(k)\triangleq col(\hat{\phi}_{y,l,(i)}(k), i=1,...,M)$
and $\hat{\phi}_{y,l,(i)}(x(k),\hat{\vartheta}_{y,l,(i)}(k))\triangleq \hat{\vartheta}_{y,l,(i)}(k)^Tg^l_{(i)}(x(k))$. 

As each $l$--th FIE fault function matches the specific fault in $\digamma_1$, it is possible to compute a dynamic threshold $\bar{r}_{y,l}$ such that $|r_{y,l,(i)}(k)| \leq \bar{r}_{y,l,(i)}(k)$ under the hypothesis that the actual fault is the $l$--th one, for each $i$ and $k$. The thresholds for actuator faults are designed as
\begin{equation}
\begin{gathered}
\bar{r}_{y,l,(i)}(k)=\bar{\eta}_{y,(i)}(k)+\alpha_{(i)} 
\left\{ 
\sum_{h=k_d}^{k-1}\right. 
\delta_{(i)}^{k-1-h}\{[\bar{\Delta} { \rho}(h)+ \\
\bar{\eta}_x(h)]+ 
\|g^l_{(i)}(h)\|[\kappa^l(h)+\|\hat{\vartheta}^l_{(i)}(h)\|\bar{b}^{-(h-k_d)}]
\}+
\left. \delta_{(i)}^{k-k_d}\bar{\epsilon}^l_x(k_d)
\right\} \, ,
\end{gathered}
\label{eq:threshold for FIE}
\end{equation}
where $\kappa^l$ is the maximum possible parameter estimation error 
\begin{equation}
\begin{gathered}
\kappa^l \triangleq \max_{\vartheta_{x,l}\in \hat{\Theta}^l}(\|\vartheta_{x,l}-\hat{\vartheta}_{x,l}(k)\|) \, .
\end{gathered}
\label{eq:kappa}
\end{equation}
The threshold derivation for sensor faults is similar and will not be repeated here. 

It is worth noting that the threshold equation of \eqref{eq:threshold for FIE} guarantees that if the $s$--th fault $\phi_{x,s}\in \digamma_1$ occurs and is learned by its FIE successfully, that residual is guaranteed to be bounded by its threshold. If at least one component of all the other FIE residuals exceeds {its} corresponding threshold, then the $s$--th fault will be successfully isolated.

\subsection{Signal-based fault diagnosis for the mooring lines}
Detecting and isolating faults in $\digamma_2$ which are associated with the mooring system may prove to be difficult using the model-based approach of previous subsection. In particular, it can be verified that for realistic values of the FOWT and mooring lines parameters, such faults do not {fulfil} the detectability and isolability conditions stated in Theorem 3.1 and Theorem 3.2 of Ferrari \textit{et al}. \cite{Ferrari-2008}.
To cope with the faults in the mooring system, a signal-based scheme is proposed, which takes advantage of the noticeable influence that such faults have on the spectrum of the tower-top acceleration ${\ddot{\alpha}}$. 
\begin{rem}\label{rem:benefit}
{Although the mooring line faults can be identified from the platform motions or mooring line tensions, the measurement of these signals are expensive. Furthermore, the accuracy of the floater motion measurement is still questionable, as it should be further validated \cite{Yamaguchi_2016}.

In comparison, the measurement of the tower-top vibration is a proven technique, which has been widely used in wind farms.
Based on the tower-top acceleration frequency analysis, two representative mooring line faults will be identified.} Such a mixed model and signal-based architecture allows to detect and isolate all the considered faults, without requiring the installation of additional, platform-motions or mooring-lines specific sensors. 
\end{rem}
In the present study, STFT \cite{Flandrin-1998} is utilized to transform the time series of the tower-top acceleration ${\ddot{\alpha}}$ as
\begin{equation}
S_{\ddot{\alpha}}(\tau,w)=\sum_{k=-\infty}^{+\infty}\ddot{\alpha}(k)h(k-\tau)e^{-jwk} \, ,
\color{black}
\label{eq:STFT}
\end{equation}

where ${S_{\ddot{\alpha}}}$ is the time-frequency spectrum of ${\ddot{\alpha}}$, while $\tau$ and $w$ denote, respectively, the transformed time and angular frequency coordinates. Furthermore, $h(k)$ is a window determining the time and frequency resolution.

Using either experimental or simulation data, several training spectra ${S_{\ddot{\alpha},q}^l(\tau,w)}$ are obtained, with $q=1,\dots,Q$. The quantity $Q$ indicates the total number of training spectra for each hypothesis, the possible hypotheses being either healthy conditions ($l=0$), a fairlead fault ($l = 10$) or an anchor fault ($l = 11$).
In order to detect and isolate mooring lines faults, the KNN algorithm \cite{Cover-1953} is run on the distance between the training spectra ${S_{\ddot{\alpha},q}^l(\tau,w)}$ and the actual one ${S_{\ddot{\alpha}}(\tau,w)}$ as
\begin{equation}
D_{K,q}^l(\tau)=\sqrt{\sum_{p=1}^{P} \left(S_{\ddot{\alpha}}(\tau,w_p)-S_{\ddot{\alpha},q}^l(\tau,w_p) \right)^2} \, ,
\color{black}
\label{eq:KNN}
\end{equation}

where $w_p$ with $p = 1, \dots, P$ denotes one frequency bin\footnote{As the STFT is computed using the discrete fast Fourier transform, only a discrete number of frequency bins are available.}.
After selecting the neighborhood size $K$, faults can be detected and isolated by the majority voting criterion \cite{Ghosh-2005}, where the widely-used rule $K\le\sqrt{Q}$ is employed following \cite{Ghosh-2005}.

%% file: sections/4_sims.tex
The effectiveness of the proposed mixed FD architecture is verified via a case study based on the newly developed 10MW FOWT benchmark. Following the case study, the {comparisons} between the developed architecture and other two classical approaches reported in literature {are} carried out. The advantages and limitations of the proposed mixed FD architecture are discussed based on the comparison {results}.

\subsection{Model configuration}
The aero-hydro-structural dynamics of the 10MW FOWT benchmark are simulated by a modified version of \emph{FAST v8.16} \cite{Jonkman-2005}. It is noted that the open-sourced code of FAST is extended by the authors to incorporate all the faults defined in Table~\ref{table:fault_function}. Such a customized FAST code is thus embedded in Simulink and connected to external blocks implementing the basic controllers, actuators and the turbine's sensors, as shown in the lower part of Fig.~\ref{Pic_block}. 

{In total, seven} Load Cases (LCs) considering laminar and turbulent wind conditions and irregular waves are simulated in the 10MW FOWT benchmark, as shown in Table~\ref{table:load_case}. 
In the first three LCs, a constant wind speed ($U_m$) of, respectively, 12m/s, 16m/s and 20 m/s is assumed at the hub height. 
Regarding LC4-LC6, a three-dimensional turbulent varying wind field, which is produced by the TurbSim\footnote{TurbSim: a stochastic inflow turbulence tool to generate realistic turbulent wind fields. https://nwtc.nrel.gov/TurbSim.} tool, is considered. {In addition to these cases above the rated wind speed (11.4m/s), a wind condition of 8m/s which is below the rated value, is considered in LC7.}
In details, it can be described as,
\begin{equation}
U_w(k) = U_m+ U_{f}(k)\, ,
\label{eq:turbSim}
\end{equation}
where the constant wind speed $U_m$ is used to specify the normal wind profile and $U_f$ is a fluctuating wind component calculated from the Normal Turbulence Model (NTM) {\cite{Ishihara_2012}}. NTM is based on the Kaimal turbulence model according to the turbulence intensities specified in the IEC with class C.

In addition, the wave conditions are computed using the JONSWAP wave spectrum model. 
{The significant wave height $H_{\mbox{s}}$ and peak-spectral period $T_{\mbox{p}}$ in all LCs are predicted by the conditional probabilistic distribution derived from the long-term observation campaigns in the North Sea \cite{Li-2015}.}
In each LC, the simulation lasts 1000s at a fixed discrete time step of $T_s =$ 0.01s.
In total, 11 faulty scenarios are simulated during each LC, as described in Table~\ref{table:fault}. In particular, the magnitude of $f_1-f_8$ are extracted from the literature \cite{Odgaard-2013, Badihi-2015}. 

\begin{table}
\setlength{\tabcolsep}{3mm}{
\caption{Fault scenarios in the case study.\label{table:fault}}
\begin{tabular}{clll}
\hline 
Number  & Description  & Parameter  &  Time \\ \hline
\textbf{Fault case A} \\ 
$f_1$  & Generator speed           & 0.95         & 
{$210\,\mbox{s}-235\,\mbox{s}$}  \\
$f_2$  & Generator power           & 1.1          & 
{$305\,\mbox{s}-330\,\mbox{s}$}  \\
$f_3$  & Blade root bending moment & {$10^4$\,kN m}  & {$400\,\mbox{s}-425\,\mbox{s}$}  \\
$f_4$  & Rotor speed               & 1.1         & 
{$495\,\mbox{s}-520\,\mbox{s}$}  \\
$f_5$  & Pitch actuator            & {0.2\,rad}     & {$590\,\mbox{s}-615\,\mbox{s}$}  \\
$f_6$  & Torque actuator           & {$20$\,kN m}   & {$685\,\mbox{s}-710\,\mbox{s}$}  \\
$f_7$  & Pitch sensor              & {0.2\,rad}      & {$780\,\mbox{s}-805\,\mbox{s}$}  \\
$f_8$  & Torque sensor             & 0.9          & 
{$875\,\mbox{s}-900\,\mbox{s}$}  \\
$f_9$  & Rotor blade sudden fault  & $c$=0.2      & {$970\,\mbox{s}-1000\,\mbox{s}$}      \\\hline
\textbf{Fault case B} \\ 
$f_{10}$ & Mooring line (fairlead)& {0\,N}           & {$300\,\mbox{s}-1000\,\mbox{s}$}      \\\hline
\textbf{Fault case C} \\ 
$f_{11}$ & Mooring line (anchor)  & {150\,m}          & {$300\,\mbox{s}-1000\,\mbox{s}$}      \\ \hline 
\end{tabular}}
\end{table}

\begin{table}
\setlength{\tabcolsep}{4mm}{
\caption{Wind and wave conditions in seven LCs.\label{table:load_case}}
\begin{tabular}{ccccc}
\hline
LC & Wind condition & $U_m$ (m/s)     & $H_{\mbox{s}}$ (m)            & $T_{\mbox{p}}$ (s)\\ \hline
1  & 
Laminar   & 
\multirow{2}*{12} & 
\multirow{2}*{2.66}  & 
\multirow{2}*{7.42}  \\
4  & 
Turbulent    \\ \hline

2  & 
Laminar      & 
\multirow{2}*{16}  & 
\multirow{2}*{3.78} & 
\multirow{2}*{7.80} \\ 
5 & 
Turbulent \\ \hline

3  & 
Laminar     & 
\multirow{2}*{20} & 
\multirow{2}*{5.13} & 
\multirow{2}*{8.47} \\
6 & Turbulent  \\ 
\hline 

{
7} & {Laminar, below rated value} &
{8} &
{1.69} &
{7.28} \\

\hline
\end{tabular}}
\end{table}
\subsection{Model-based FD for the turbine system}
For the purpose of developing the FDAE and FIEs in the model-based FD scheme, the linear model is obtained via the subspace identification.

The normal {operating} condition of the 10MW FOWT benchmark, which is articulated in section \ref{sec:2}, is simulated via the extended FAST code to produce data in healthy conditions for subspace identification at a reduced time step of $20\times T_s =$ 0.2s. 
In order to guarantee the persistent excitation of the system dynamics, the generalized binary noise \cite{Tulleken-1990} is added into the input pitch angle and generator torque ($\theta_{\mbox{in}}$ and $T_{g,\mbox{in}}$) corresponding to steady state conditions at each of LCs.
Then, the subspace identification procedure \cite{Gestel-2001} is performed to obtain the linear model of the 10MW FOWT benchmark.

\begin{figure}
\centering \includegraphics[width=1\columnwidth]{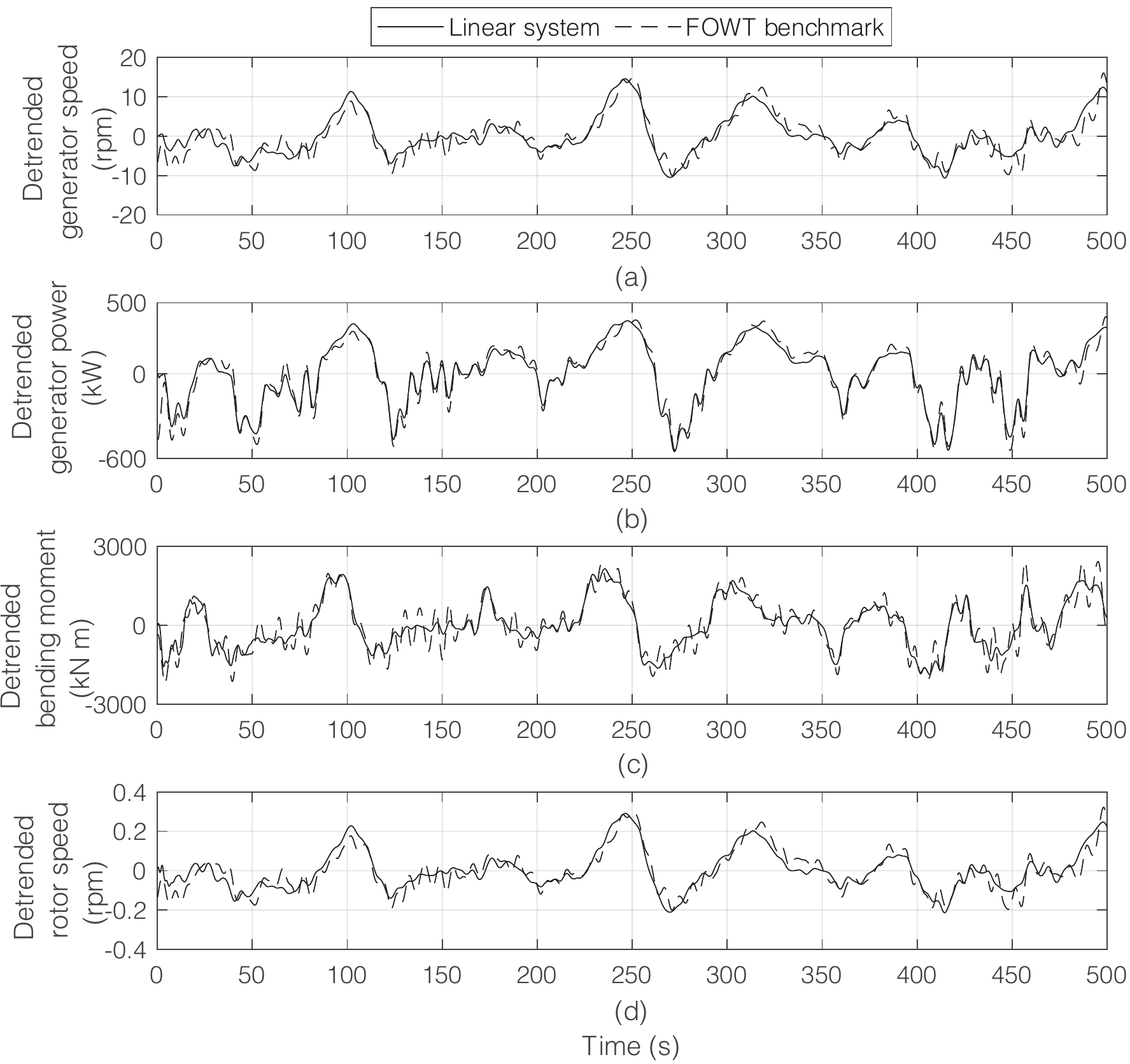}
\caption{Results of the subspace identification in LC1.(a) Generator speed, VAF: 86\%, (b) Generator power, VAF: 96\%, (c) Bending moment, VAF: 86\%, (d) Rotor speed, VAF: 86\%.}
\label{Pic_ident1} %
\end{figure}

\begin{figure}
\centering \includegraphics[width=1\columnwidth]{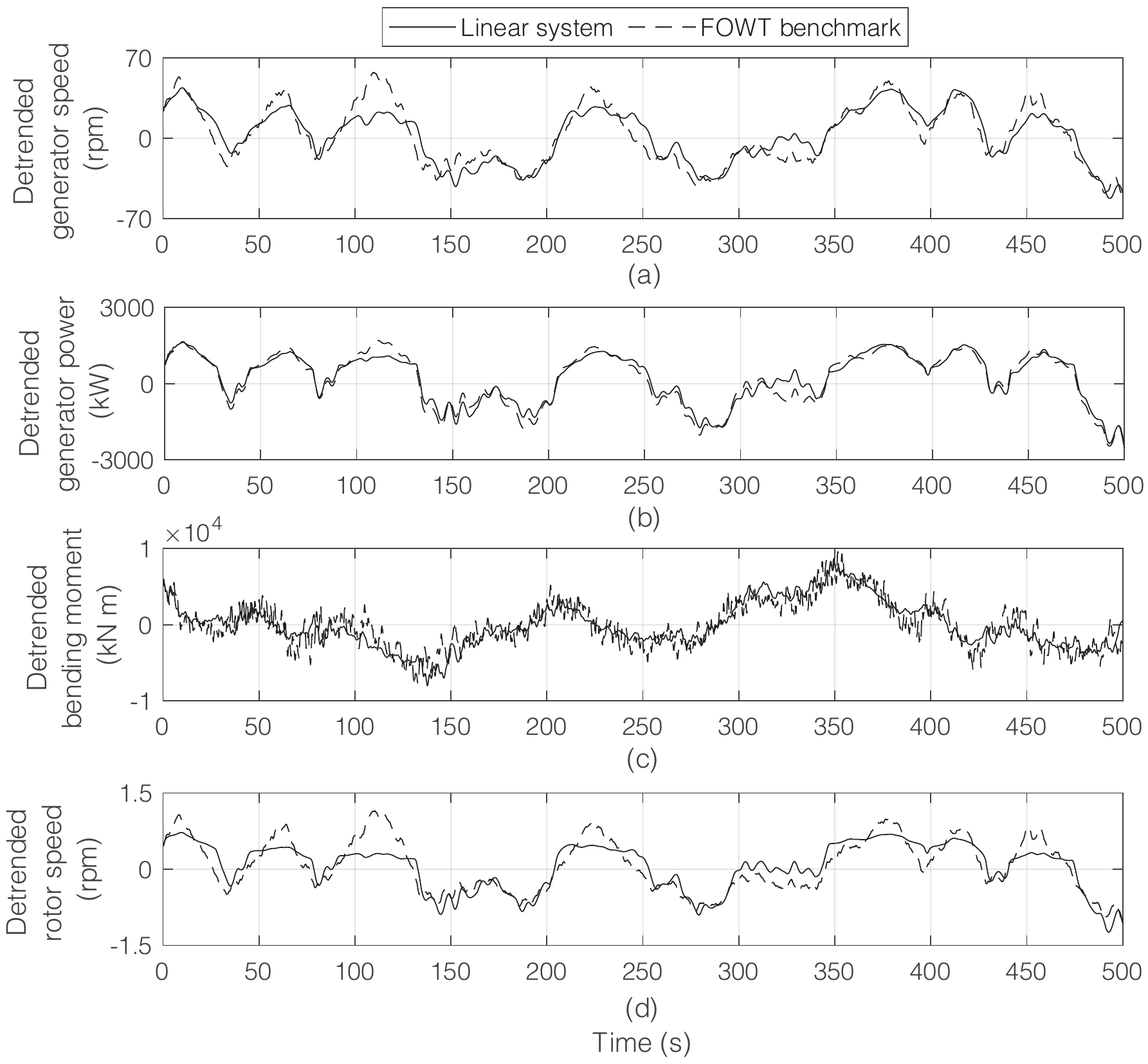}
\caption{Results of the subspace identification in LC4.(a) Generator speed, VAF: 77\%, (b) Generator power, VAF: 96\%, (c) Bending moment, VAF: 74\%, (d) Rotor speed, VAF: 75\%.}
\label{Pic_ident_tur12} %
\end{figure}

Figs.~\ref{Pic_ident1}-\ref{Pic_ident_tur12} shows the results of the subspace identification run in laminar wind condition (LC1) and turbulent wind condition (LC4). The performance of the subspace identification is evaluated by checking the Variance-Accounted-For (VAF) \cite{van-2008}. It can be seen that all the VAFs of the identified system in LC1 and LC4 are higher than 86\% and 74\%, which indicate that the predicted value from the linear system is an acceptable match for the real signals.
Compared to LC1, it shows that the VAFs of the generator speed, bending moment and rotor speed in LC4 are lower. One plausible explanation is that the high-frequency turbulence can not be fully captured by the linearized model. Despite of such a deviation in the high frequency region, it is considered that the performance of the subspace identification is acceptable in both laminar and turbulent wind conditions. 
Other plots of the subspace identifications in LCs 2-3 and 5-{7} show similar patterns and are thereby omitted for the sake of brevity.


\begin{figure}
\centering \includegraphics[width=1\columnwidth]{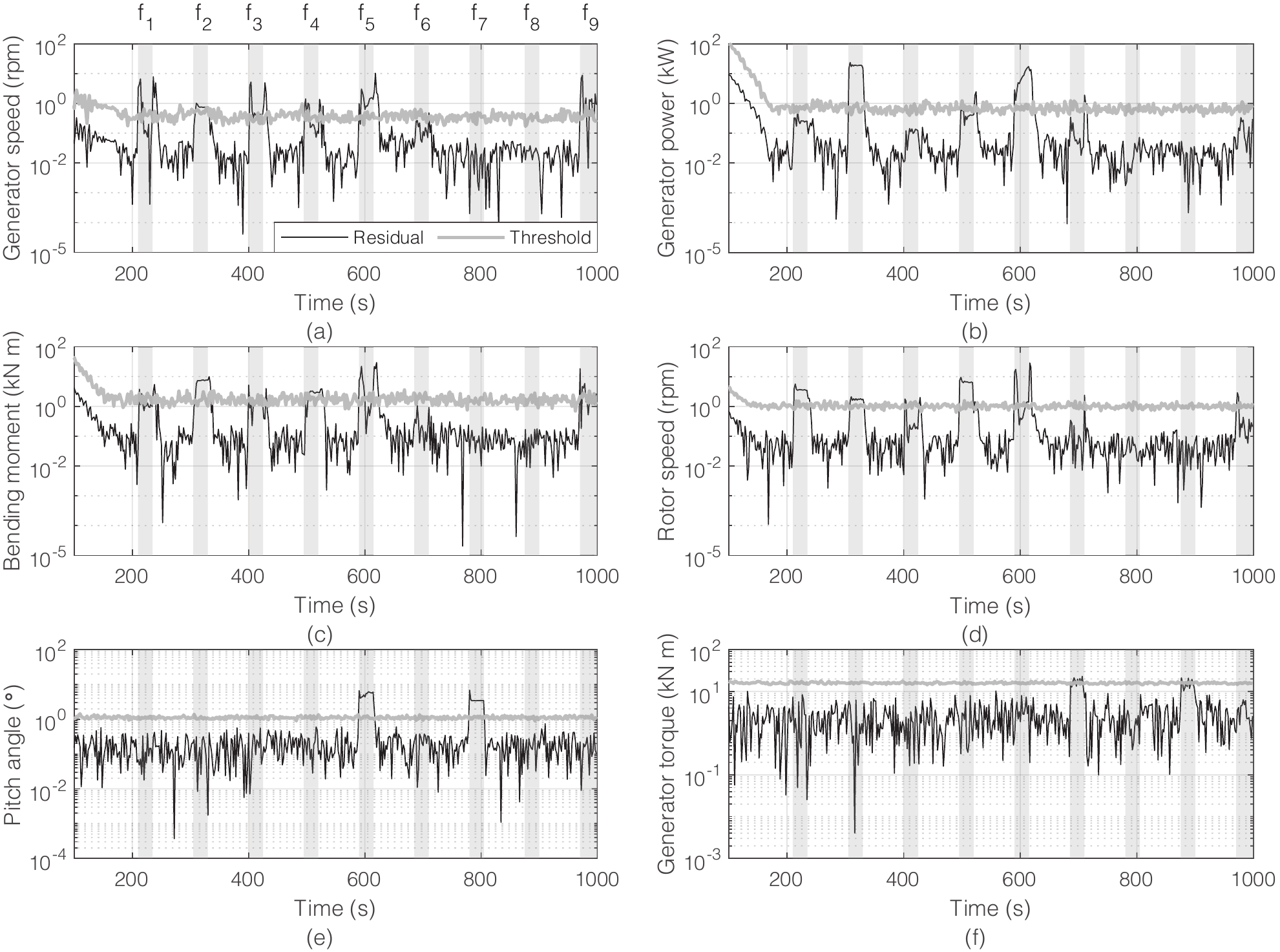}
\caption{Residual generation and model-based fault detection in LC1. Light grey shadow indicates the fault scenarios corresponding to the fault time in the Table \ref{table:fault}. (a) Generator speed, (b) Generator power, (c) Bending moment, (d) Rotor speed, (e) Pitch angle, (f) Generator torque.}
\label{Pic_model_based_detection_12} %
\end{figure}

Based on the derived linear model and on a discrete time step of $20\times T_s=$ 0.2s, the simulation results of the FDAE in LC1 are presented in Fig. \ref{Pic_model_based_detection_12} for discussions. In general, it is noticeable that the faults $f_1$-$f_9$ are detected successfully in LC1 as the residuals in faulty scenarios exceed the corresponding thresholds, which indicate the effectiveness of the proposed FDAE. 
In addition to these actuator and sensor faults, the structural fault, namely the rotor blade fault in the simulation time of {970s to 1000s}, is successfully detected by the proposed mixed FD architecture. 



\begin{figure}
\centering \includegraphics[width=1\columnwidth]{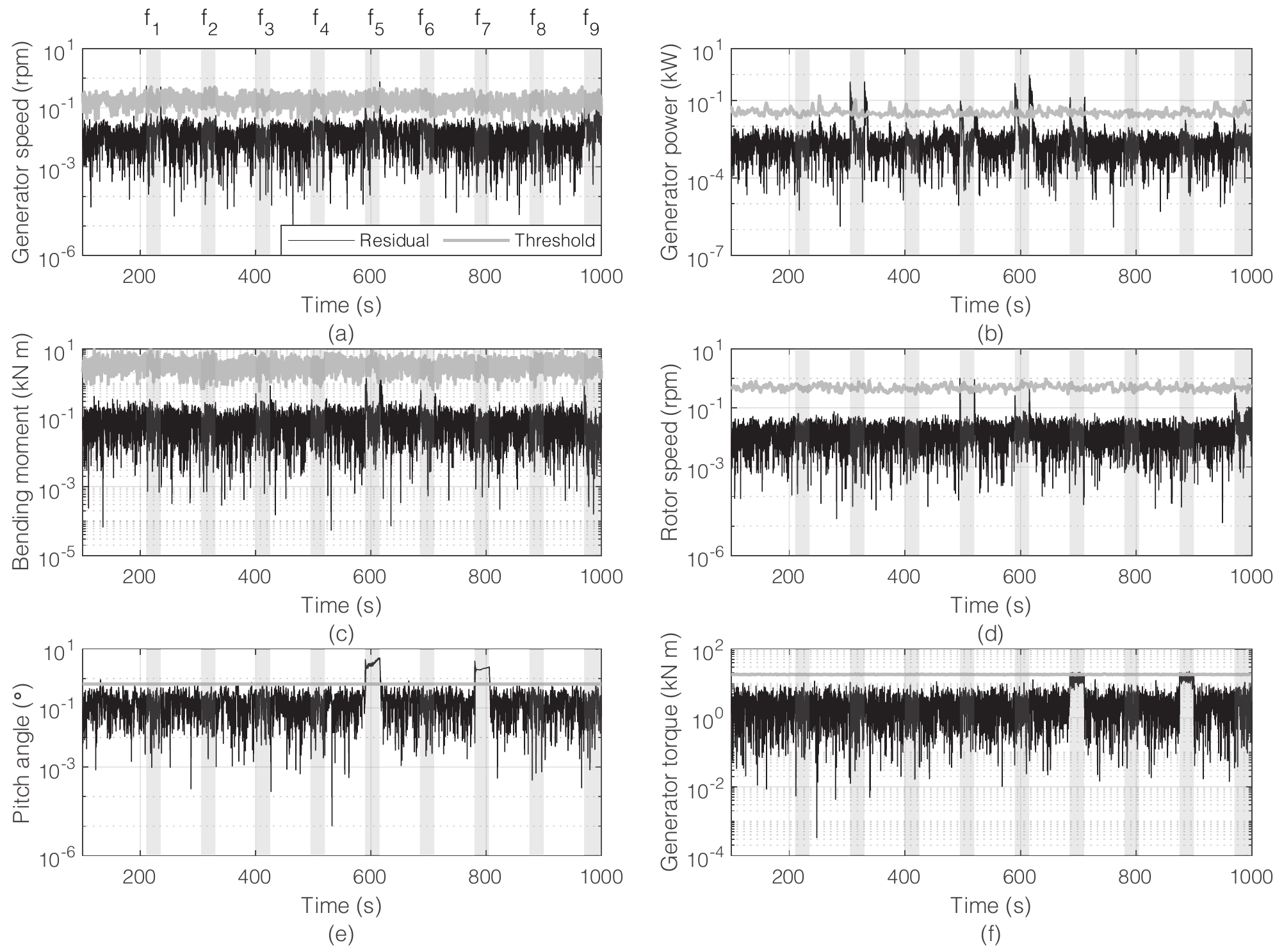}
\caption{Residual generation and model-based fault detection in LC4. Light grey shadow indicates the fault scenarios corresponding to the fault time in the Table \ref{table:fault}. (a) Generator speed, (b) Generator power, (c) Bending moment, (d) Rotor speed, (e) Pitch angle, (f) Generator torque.}
\label{Pic_fault_detection_tur12} %
\end{figure}

Compared to the laminar wind condition in LC1, a much more significant disturbance is found in the residuals of LC4 due to the turbulent wind, according to Fig. \ref{Pic_fault_detection_tur12}. In general, most of faults can be detected successfully by the proposed mixed FD architecture, except {for some detection delay for $f_3$ and $f_9$}. One plausible explanation is such a bending moment fault is buried in the fluctuations in the output signals due to the drastic turbulence and high wind speeds, and as such is too small to be detected. 

Similar results are obtained for other LCs, and are summarized in Table~\ref{table:summary}. In order to easily quantify detection robustness, {an indicator called Maximum Ratio between Residuals and Thresholds (MRRT) is introduced}. An MRRT is larger than 1 means that for at least one sample the residual exceeds the threshold:
a higher MRRT implies that the FD architecture has a better fault detection capability as it is more likely to be able to detect and isolate smaller faults.

\begin{table}

\setlength{\tabcolsep}{4.6mm}{
\caption{MRRT of the proposed FD architecture in seven LCs.\label{table:summary}}
\begin{tabular}{cccccccc}
\hline
& LC1 & LC2 & LC3 & LC4 & LC5 & LC6 & LC7\\ \hline
$f_1$ & 24.7  & 1.5  & 6.7  & 2.4  & 1.1  & 1.1   &  25.9  \\
$f_2$ & 50.5  & 7.5  & 7.9 & 20.2  & 8.1  & 1.9   &  16.7  \\
$f_3$ & 21.3  & 1.7  & 3.2  & 1.0  & 1.5  & 1.0   &  5.9  \\
$f_4$ & 11.1  & 3.2  & 12.5  & 3.4  & 1.6  & 1.2  &  50.5  \\
$f_5$ & 26.1  & 3.4  & 44.9  & 11.7  & 14.8  & 34.3 &  61.4  \\
$f_6$ & 2.7  & 9.6  & 5.0  & 4.0  & 1.8  & 1.5    &  58.9  \\
$f_7$ & 6.5  & 1.7  & 6.7   & 6.0  & 2.7  & 11.9  &  10.9  \\
$f_8$ & 1.5  & 1.8  & 4.0   & 1.2  & 1.2  & 1.2   &  0.8  \\
$f_9$ & 32.6  & 1.7  & 5.9  & 2.0  & 1.0  & 1 .0  &  7.3  \\ 
\hline 
\end{tabular}}
\end{table}

Compared to laminar wind conditions, it can be seen from Table~\ref{table:summary} that MRRT in turbulent wind conditions (LC4-LC6) is much lower than in laminar ones. 
Particularly, the MRRT in LC4 is lower than LC1 by {$\sim 49\%$}, which can be explained by the fact that the wind turbulence will cause the FOWT to operate more frequently away from the operating point around which the dynamics used by the FDAE have been linearized. {It induces significant nonlinear effects on the mixed FD architecture and reduces the fault detection capability. 
However, some values in LC4-LC6 are higher than in corresponding laminar wind conditions, such as $f_6$ in LC4, $f_2$, $f_5$ and $f_7$ in LC5 and $f_7$ in LC6.}
One plausible explanation is that in some cases the additional uncertainty due to the wind turbulence may instead positively interfere with the other components affecting the residual and make it larger, thus easing detection. 
{
Regarding LC7 below the rated wind speed, the proposed FD architecture in general shows similar results as in LC1 above the value one. 
It is interesting to note that $f_8$ is hard to be detected in this case, which shows a lower MRRT in LC7. 
The reason for this is that the generator torque in this case is smaller than the rated value. This makes the scaling sensor fault of the torque too small to be detected.
}



After a successful fault detection, a bank of FIEs is activated for the isolation step. More specifically, $8$ FIEs are employed to match the structured faults ($f_1$-$f_8$) within $\digamma_1$, while fault $f_9$, being unstructured, is learned by the general adaptive approximator of the FDAE. 

In detail, the FDAE approximator $\hat{\phi}_0$ in eq.~\eqref{eq:approximator} is enabled to approximate the unknown faults and unstructured component faults. 
It is based on a 6-input, 20-output RBF neural network with one hidden layer of $60$ fixed neurons covering all the admissible values for state and input variables.
$\hat{\vartheta}_0$ will have $20\cdot60$ components containing the weights by which the hidden layer outputs are linearly combined to calculate the network output.

\begin{figure}
\centering \includegraphics[width=1\columnwidth]{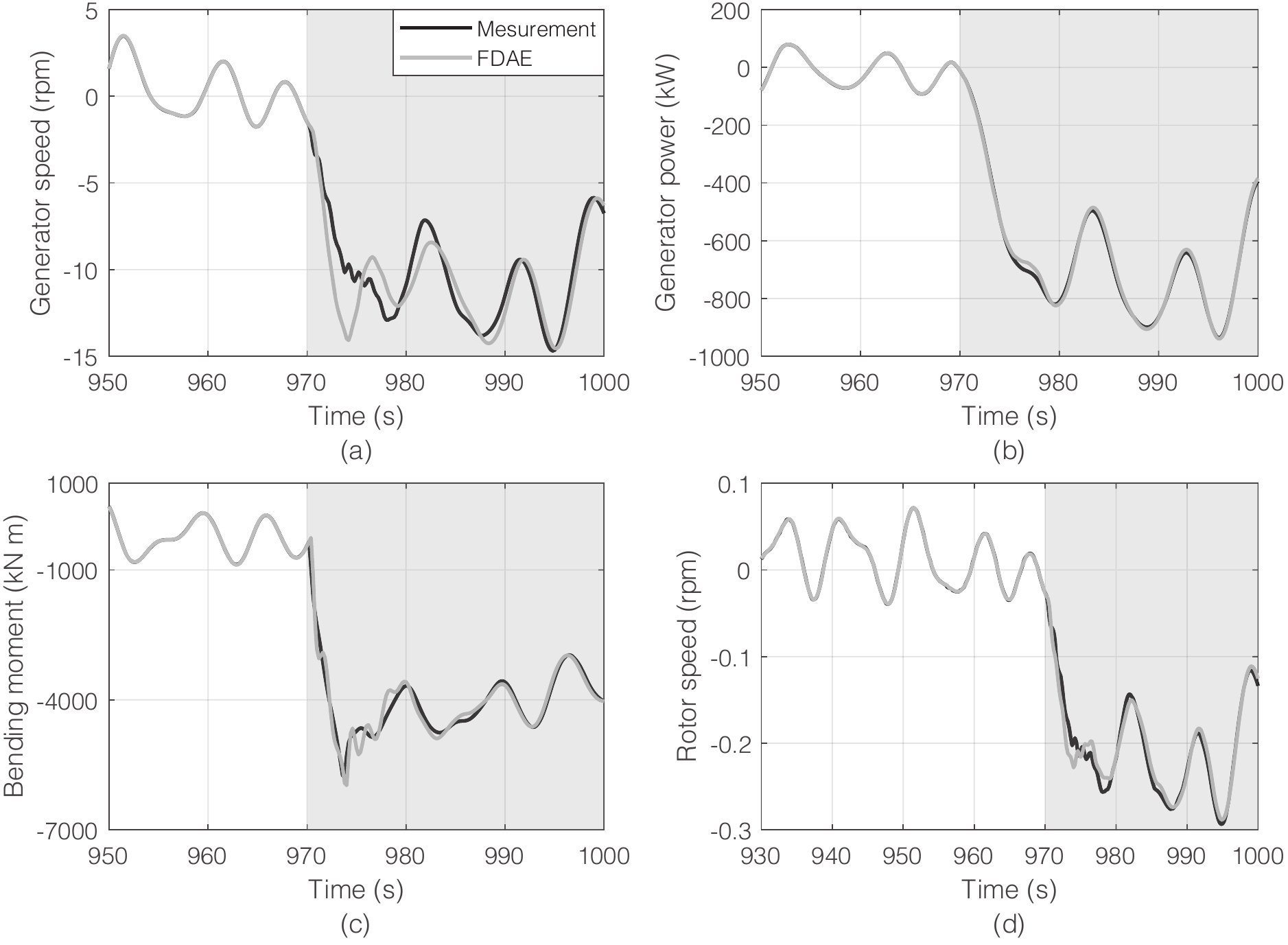}
\caption{Comparisons between measurements and FDAE outputs in LC1. Light grey shadow indicates the fault scenarios corresponding to the fault time in the Table~\ref{table:fault}. (a) Generator speed, (b) Generator power, (c) Bending moment, (d) Rotor speed.}
\label{Pic_FDAE_12} %
\end{figure}

\begin{figure}
\centering \includegraphics[width=1\columnwidth]{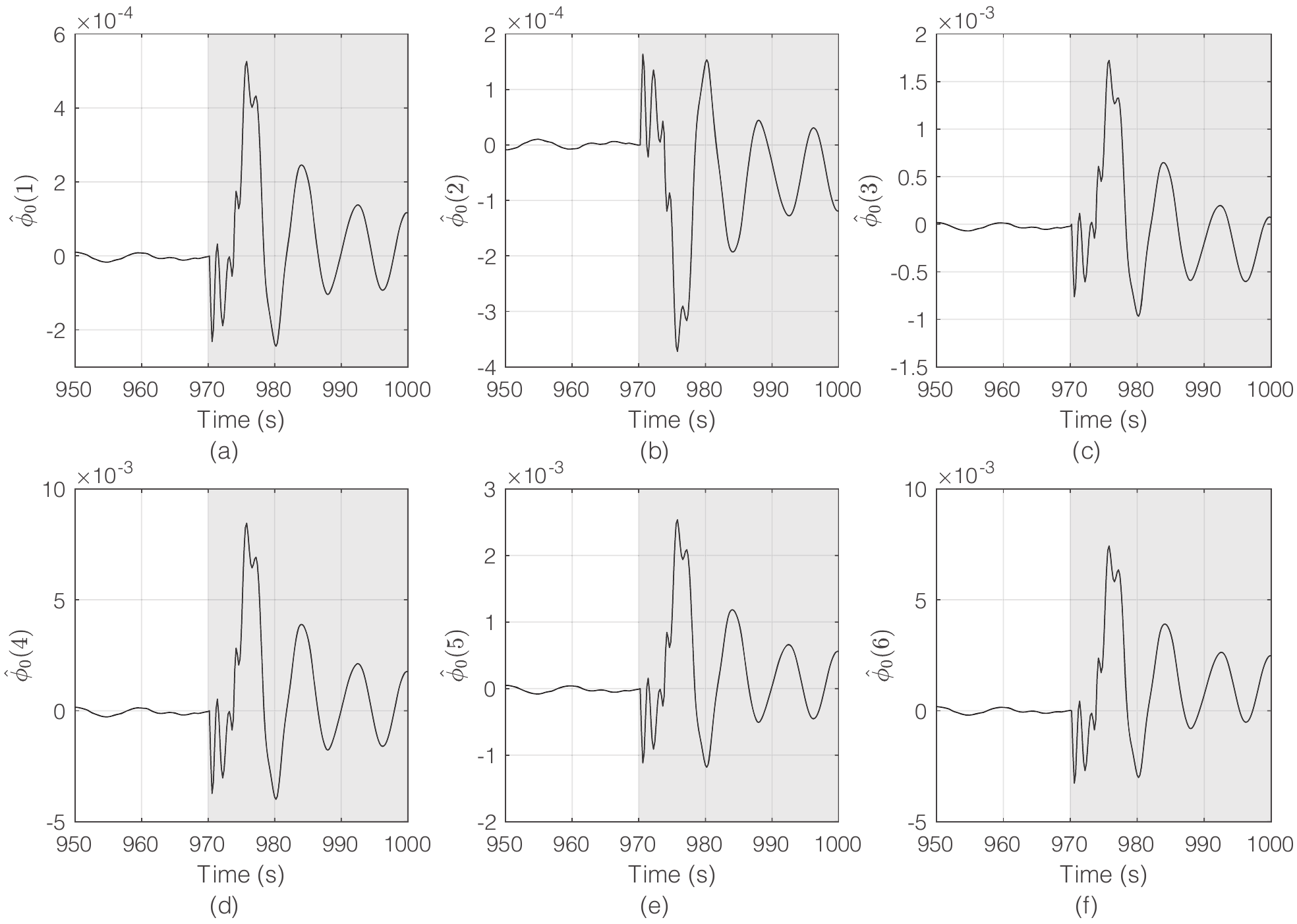}
\caption{Approximation (first 6 outputs) of the RBF neural network in LC1. Light grey shadow indicates the fault scenarios corresponding to the fault time in the Table~\ref{table:fault}. (a) $\hat{\phi}_0{(1)}$, (b) $\hat{\phi}_0{(2)}$, (c) $\hat{\phi}_0{(3)}$, (d) $\hat{\phi}_0{(4)}$, (e) $\hat{\phi}_0{(5)}$, (f) $\hat{\phi}_0{(6)}$.}
\label{Pic_FDAE_zeta_12} %
\end{figure}

By including the approximation of the neural network, the results of FDAE in LC1 is presented in Fig. \ref{Pic_FDAE_12}. It can be seen that such an unstructured blade fault in $f_9$ is eventually learned successfully by the online adaptive approximator $\hat{\phi}_0$. Anyway, some deviations between the measurements and FDAE are observed during {970s-990s} due to the oscillating behaviour of the neural network output, which is induced by the learning law and partially illustrated in Fig. \ref{Pic_FDAE_zeta_12}.

\begin{figure}
\centering \includegraphics[width=1\columnwidth]{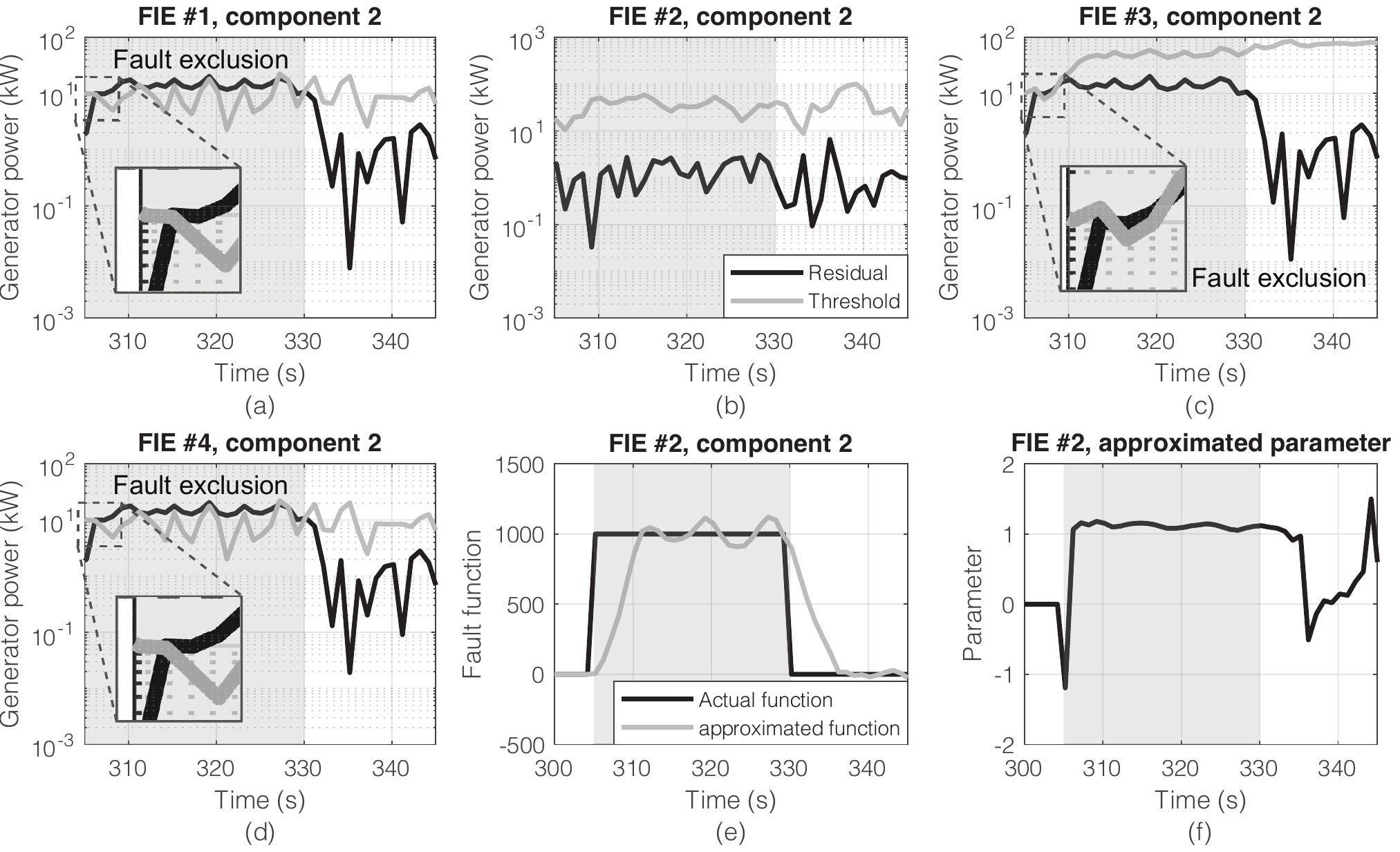}
\caption{Fault isolation for the sensor fault $(f_2)$ in LC2. Light grey shadow indicates the fault scenarios corresponding to the fault time in the Table~ \ref{table:fault}. (a) FIE $\#$2, component 1, (b) FIE $\#$3, component1, (c) FIE $\#$4, component 1, (d) FIE $\#$5, component 1, (e) Approximated fault function, (f) Approximated parameter.}
\label{Pic_isolation_f2} %
\end{figure}

In addition to $f_9$, other faults ($f_1-f_8$) are isolated by the designed FIEs.
Fig. \ref{Pic_isolation_f2} shows the results of the fault isolation and the approximated fault function of the sensor fault ($f_2$) in LC2. 
It is worth nothing that $f_2$ is isolated at the {early stage} of the fault scenario ({306s}), since only the residuals produced by the FIE $\#$2 are within the range specified by the corresponding thresholds. The residuals from other FIEs($\#$1, $\#$3 and $\#$4) exceed their thresholds due to the mismatch between these FIEs and the fault function. After {330s}, the residuals from FIEs ($\#$1, $\#$3 and $\#$4) reduce gradually due to the recovered nominal healthy condition of the FOWT. In addition, the actual fault function of $f_2$, which essentially is a multiplicative fault as shown in Table~\ref{table:fault_function}, is successfully approximated according to Fig. \ref{Pic_isolation_f2}(e-f). 
It should be noted that FIE $\#$2 slightly underestimates the actual fault function at the beginning ({305s-311s}) according to Fig. \ref{Pic_isolation_f2}(e). After that, the real value of the fault is gradually approximated by FIE $\#$2. 

\begin{figure}
\centering \includegraphics[width=1\columnwidth]{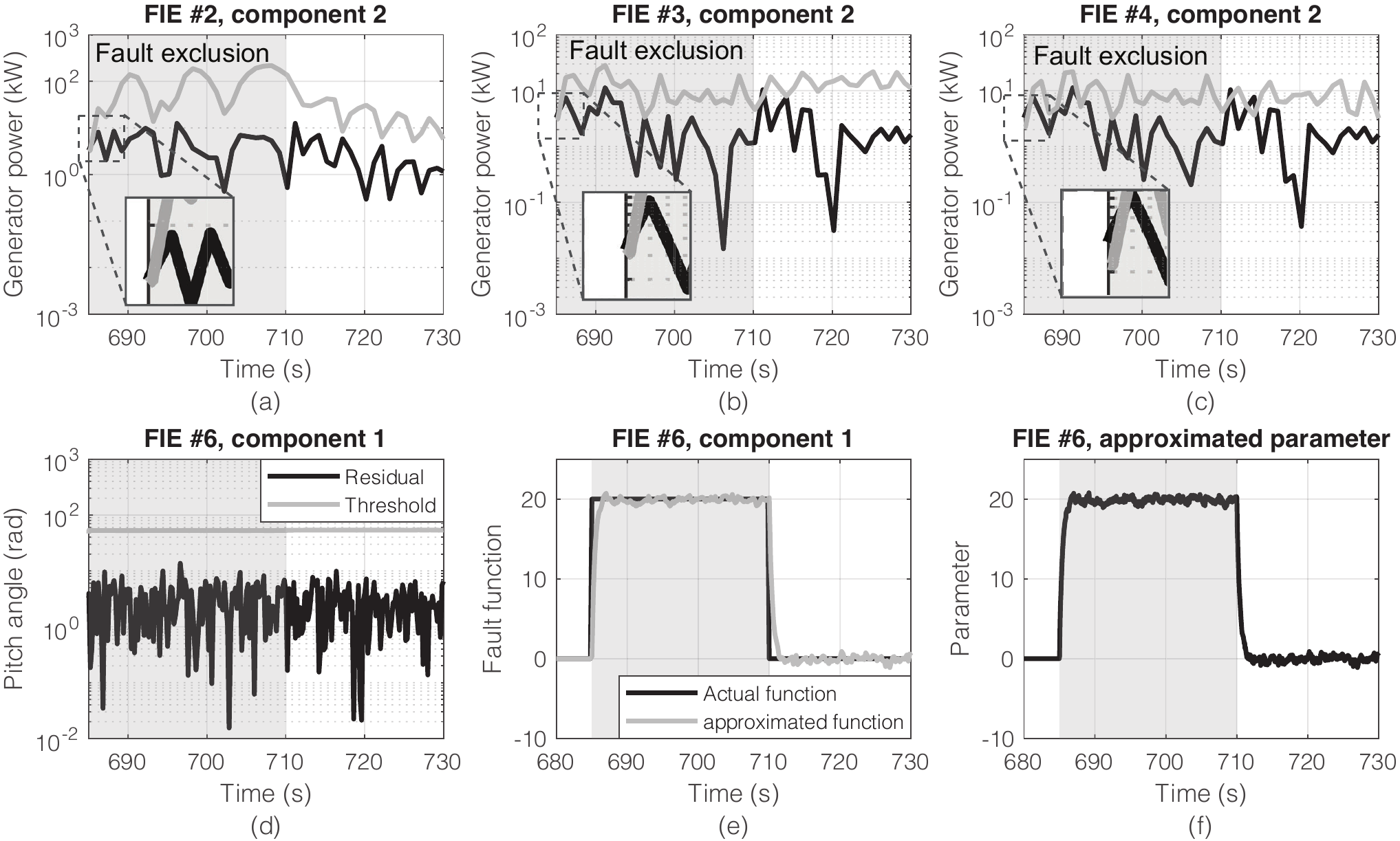}
\caption{Fault isolation for the actuator fault $(f_6)$ in LC2. Light grey shadow indicates the fault scenarios corresponding to the fault time in the Table~ \ref{table:fault}. (a) FIE $\#$2, component 2, (b) FIE $\#$3, component 2, (c) FIE $\#$4, component 2, (d) FIE $\#$6, component 1, (e) Approximated fault function, (f) Approximated parameter.}
\label{Pic_isolation_f6} %
\end{figure}

\begin{figure}
\centering \includegraphics[width=1\columnwidth]{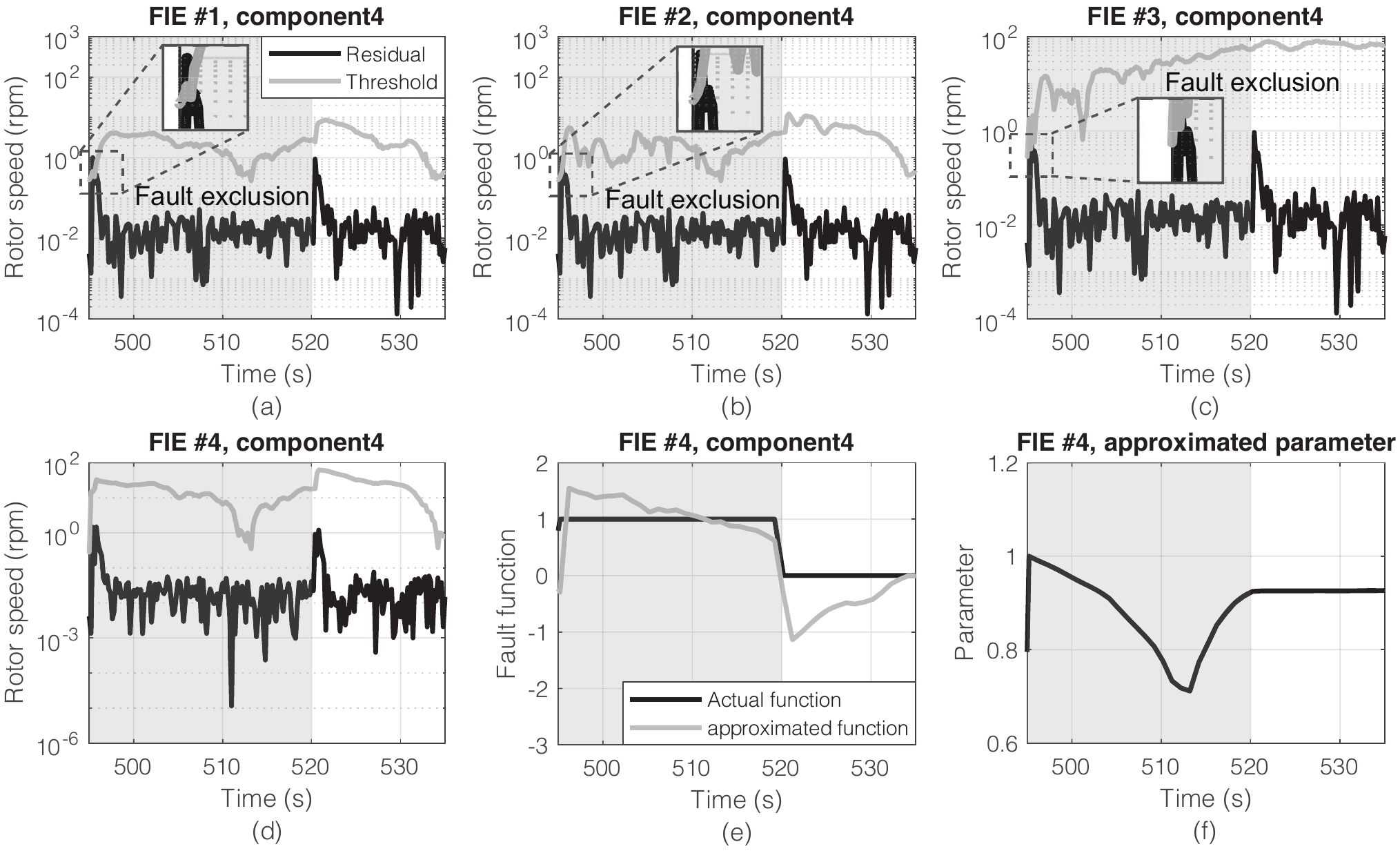}
\caption{Fault isolation for the actuator fault $(f_4)$ in LC4. Light grey shadow indicates the fault scenarios corresponding to the fault time in the Table~ \ref{table:fault}. (a) FIE $\#$2, component 4, (b) FIE $\#$3, component 4, (c) FIE $\#$4, component 4, (d) FIE $\#$6, component 4, (e) Approximated fault function, (f) Approximated parameter.}
\label{Pic_isolation_f4_tur} %
\end{figure}

Similarly, the isolation results of the actuator fault ($f_6$) in LC2 and the sensor fault ($f_4$) in LC4 are presented in Figs. \ref{Pic_isolation_f6}-\ref{Pic_isolation_f4_tur}. 
It is concluded that both the additive fault function of the generator torque and  multiplicative fault function of the rotor speed, are gradually learned by FIE$\#6$ and FIE$\#4$. This verifies the effectiveness of the fault isolation capability of the proposed mixed FD architecture.

\subsection{Signal-based FD for the mooring lines}
Regarding the signal-based FD for the two representative mooring line faults in $\digamma_2$, the Hamming window function \cite{Bouchikhi-2011} is used for computing STFT and obtain the training spectral pattern of ${\ddot{\alpha}}$ in healthy and faulty conditions.
The length of the moving Hamming window is 1000 samples while the overlap between consecutive windows is 10 samples. 
The length of the fast Fourier transform, on the other hand, is 2000 samples. 
During the KNN classification, $P$, $Q$ and $K$ in eq.~\eqref{eq:KNN} are set to be 1000, 64 and 8, respectively. 

\begin{figure}
\centering \includegraphics[width=0.8\columnwidth]{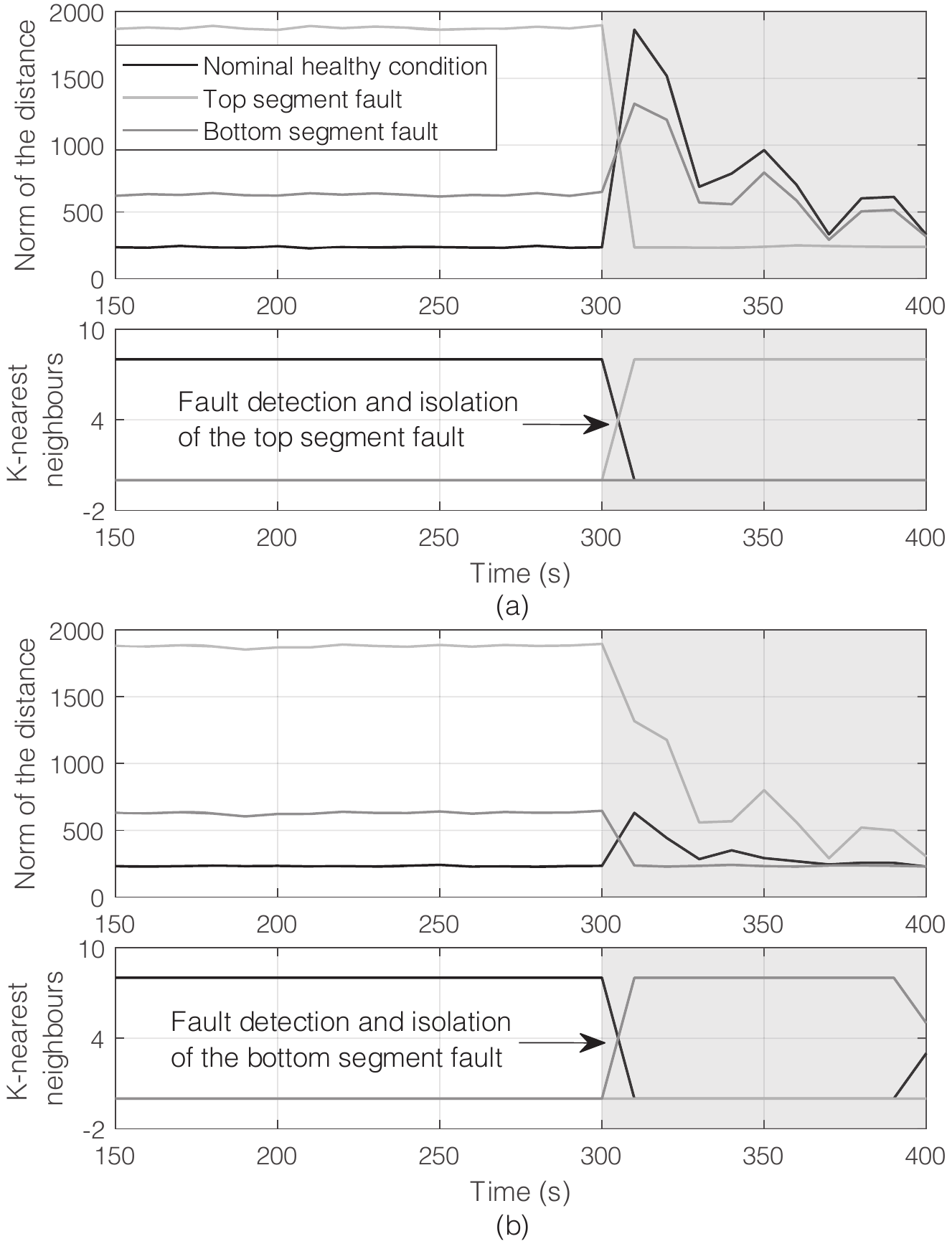}
\caption{Norm of the distance between the spectral patterns and the measurement output and KNN for the signal-based FD of the mooring line faults in LC3 ($K=8$). Light grey shadow indicates the fault scenarios corresponding to the fault time in the Table~ \ref{table:fault}. (a) Top segment fault. (b) Bottom segment fault.}
\label{Pic_mooring} %
\end{figure}

\begin{figure}
\centering \includegraphics[width=0.8\columnwidth]{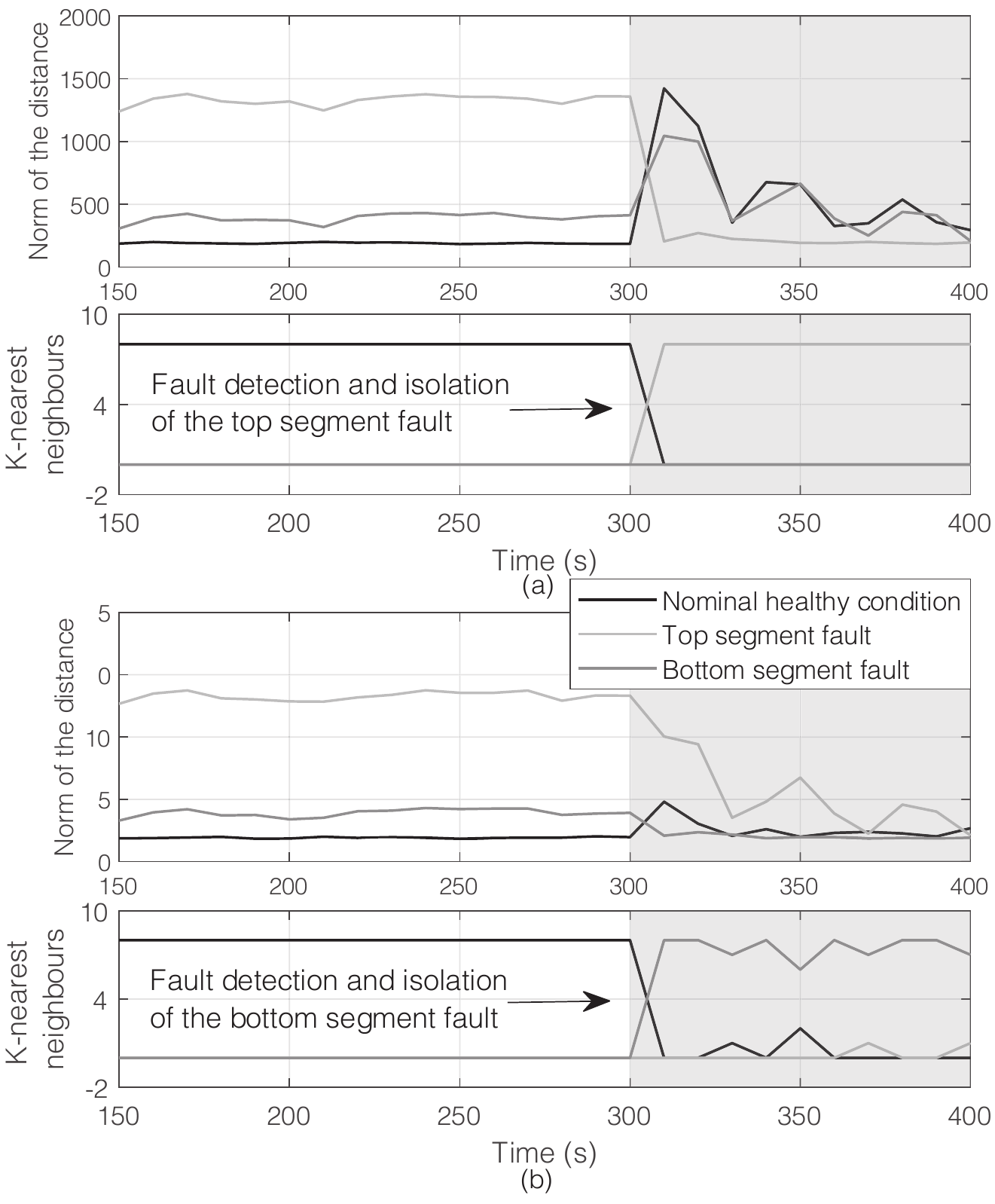}
\caption{Norm of the distance between the spectral patterns and the measurement output and KNN for the signal-based FD of the mooring line faults in LC4 ($K=8$). Light grey shadow indicates the fault scenarios corresponding to the fault time in the Table~ \ref{table:fault}. (a) Top segment fault. (b) Bottom segment fault.}
\label{Pic_mooring_12_tur} %
\end{figure}

Based on eqs.~\eqref{eq:STFT}-\eqref{eq:KNN}, some selected FD results of the mooring line faults are presented in Figs.~\ref{Pic_mooring}-\ref{Pic_mooring_12_tur} for discussion.
In general, KNNs derived from the distance between the spectral patterns and the measurement output are able to indicate the state of the mooring lines. 
Both mooring line faults are successfully detected and isolated at around 310s when the neighbors of the faulty signals indicate the FOWT benchmark violating nominal healthy conditions. 
Moreover, it is evident from {Figs.~\ref{Pic_mooring}-\ref{Pic_mooring_12_tur}} that the detection delay time \cite{Ding-2008} is around 10s, which is dependent on the moving Hamming window length utilized in the Fourier transform. Considering the time windows utilized in the transform ($1000\times T_s$), such a detection time is acceptable. 
It is worth noting that the FD results in the turbulent wind condition (LC4) show similar patterns as the laminar one (LC1), which indicates the robustness of the signal-based scheme in the proposed mixed FD architecture.

\subsection{Comparison to other approaches}
In order to illustrate the advantages and limitations of the proposed mixed FD architecture, two classical approaches, which have been developed by previous researchers, are implemented for the 10MW FOWT benchmark reported in the present study. Briefly, both of them are introduced here. 

\begin{enumerate}
\item Principal Component Analysis (PCA): this well--known technique employs an orthogonal transformation to convert a series of correlated variables into linearly uncorrelated variables. 
Kr\"uger \textit{et al.} \cite{Kruger_pcafda}, Pozo \textit{et al.} \cite{Francesc_pca} and Wang \textit{et al.} \cite{Wang_pca} employed such a PCA-based approach to detect and identify faults in wind turbines.

\item Dynamic Principal Component Analysis (DPCA): The DPCA approach uses the time-lagged version of the system input and output to develop a model for monitoring purposes, and has been widely used for the fault detection of dynamic systems \cite{Kruger_pcafda, Wang_pca}.
Rato and Reis \cite{Rato_2013} established the DPCA model based on decorrelated residuals to detect several faults and illustrated the reliability of such an approach in the fault detection. 
\end{enumerate}

In order to provide detection data for comparisons, both of aforementioned two classical approaches, PCA \cite{Wang_pca} and DPCA \cite{Rato_2013} are implemented and applied to detect faults in the 10MW FOWT benchmark.
{Comparisons} between the mixed FD architecture developed in the present study and these two classical approaches {are} carried out. The FD results are summarized in Table~\ref{table:compare}. 

\begin{table}

\setlength{\tabcolsep}{1.4mm}{
\caption{Fault detection results of three architectures, where $\surd$ is accurate detection, $\circ$ is delay detection with more than one sample, $\times$ is missed detection). \label{table:compare}}
\begin{tabular}{cccccccccccccccccccccc} 
\hline 
\multirow{2}*{LC}  & \multicolumn{7}{c}{Mixed FD}  & \multicolumn{7}{c}{PCA} &  \multicolumn{7}{c}{DPCA}  \\ 
 & 1 & 2 & 3 & 4 & 5 & 6 & {7} 
 & 1 & 2 & 3 &4 & 5 & 6 & {7}
 & 1 & 2 & 3 &4 & 5 & 6 & {7}\\ \hline
$f_1$ & $\surd$ & $\circ$ & $\surd$  &  
$\surd$ & $\circ$ & {$\circ$}   &  
{$\surd$}      & 
$\surd$ & $\circ$  & $\times$  & 
$\times$ & $\times$ & $\times$   & 
{$\surd$}       &
$\surd$  & $\circ$ & $\times$  & 
$\times$ & $\times$ & $\times$  &
{$\times$}       \\

$f_2$ & $\surd$ & $\surd$ & $\surd$  & 
$\surd$ & $\surd$ & $\surd$   &
{$\surd$}       &
$\surd$ & $\surd$  & $\circ$    & 
$\surd$ & $\surd$ & $\surd$   &
{$\times$}       &
$\surd$  & $\circ$  & $\surd$   &
$\times$ & $\surd$ &$\surd$  &
{$\times$}       \\

$f_3$ & $\surd$ & {$\circ$} & $\surd$   &  
{$\circ$} & {$\circ$} & {$\circ$}         & 
{$\surd$}       &
$\surd$ & $\surd$  & $\surd$           & 
$\times$ & $\circ$ & $\times$          &
{$\surd$}       &
$\surd$  & $\surd$ & $\surd$           & 
$\times$ & $\circ$ &$\times$ &
{$\surd$}       \\ 

$f_4$ & $\surd$ & $\surd$ & $\surd$    & 
$\surd$ & $\surd$ & $\circ$  &
{$\surd$}       &
$\surd$ & $\surd$  & $\surd$ & 
$\times$ & $\circ$  & $\times$ &
{$\surd$}       &
$\surd$ & $\surd$ & $\circ$ & 
$\times$ & $\surd$ & $\times$ & 
{$\surd$}      \\

$f_5$ & $\surd$ & $\surd$ & $\surd$    & 
$\surd$ & $\surd$ & $\surd$  & 
{$\surd$}       &
$\surd$ & $\circ$  & $\circ$   & 
$\circ$  & $\times$ &$\surd$  & 
{$\surd$}       &
$\times$ & $\surd$ & $\surd$ & 
$\circ$  & $\times$ &$\surd$ &
{$\surd$}      \\

$f_6$ & $\surd$ & $\surd$ & $\surd$   & 
$\surd$ & $\surd$ & $\circ$  & 
{$\surd$}       &
$\surd$ & $\circ$  & $\circ$  & 
$\surd$  & $\surd$ &$\surd$  & 
{$\surd$}       &
$\surd$  & $\surd$ & $\surd$  & 
$\times$ & $\surd$ &$\circ$  &
{$\surd$}     \\

$f_7$ & $\surd$ & $\surd$ & $\surd$   & 
$\surd$ & $\surd$ & $\surd$  &
{$\surd$}       &
$\surd$ & $\surd$  & $\surd$  & 
$\surd$ & $\times$ &$\surd$  &
{$\surd$}       &
$\surd$ & $\circ$ & $\surd$ & 
$\times$ & $\times$ &$\times$ &
{$\surd$}     \\

$f_8$ & $\surd$ & $\circ$ & $\circ$   & 
$\surd$ & $\circ$ & $\circ$  &
{$\times$}       &
$\surd$ & $\surd$  & $\surd$   & 
$\surd$  & $\surd$ &$\surd$  &
{$\surd$}       &
$\surd$  & $\circ$ & $\surd$   & 
$\times$ & $\surd$ &$\circ$  &
{$\times$}  \\

$f_9$ & $\surd$ & $\surd$ & $\surd$  & 
$\circ$ & $\surd$ & {$\circ$}  &
{$\surd$}       &
$\surd$ & $\surd$  & $\surd$   
& $\circ$ & $\times$ &$\times$  &
{$\surd$}       &
{$\circ$}  & {$\times$} & $\surd$ & 
$\circ$ & $\times$ &$\times$  &
{$\surd$}      \\

$f_{10}$ & $\circ$ & $\circ$ & $\circ$  & 
$\circ$ & $\circ$ & $\circ$  &
{$\circ$}       &
$\times$  & $\times$  & $\times$  &
$\times$ & $\times$ &$\times$  &
{$\times$}       &
$\times$   & $\times$ & $\times$ &
$\times$ & $\times$ &$\times$ &
{$\times$}  \\

$f_{11}$ & $\circ$ & $\circ$ & $\circ$  &
$\circ$ & $\circ$ & $\circ$  &
{$\circ$}       &
$\times$  & $\times$ & $\times$ &
$\times$  & $\times$ & $\times$  &
{$\times$}       &
$\times$   & $\times$ & $\times$ &
$\times$ & $\times$ & $\times$ &
{$\times$}   \\

\hline
\end{tabular}}
\end{table}

{It is discerned that the proposed mixed FD architecture detected most expected faults in laminar wind conditions (LC1-LC3 and LC7), except for some detection delay on $f_1$, $f_3$, $f_8$, $f_{10}$ and $f_{11}$ and missed detection on $f_8$ in the case below the rated wind speed. However, the classical PCA and DPCA have missed detection in LC2, LC3 and LC7 several times.}
{Under turbulent wind conditions (LC4-LC6), the mixed FD architecture had more detection delay for $f_1$, $f_3-f_4$, $f_6$, $f_8-f_{11}$.}
{PCA failed to detect $f_1$, $f_3-f_4$, $f_{10}-f_{11}$ and was not always able to detect $f_9$. DPCA instead failed to detect $f_1$, $f_3-f_4$, $f_7$, $f_{10}-f_{11}$ and was inconsistent in detecting $f_2$, $f_6$ and $f_8-f_9$, thus seeming the less reliable method for detecting faults in turbulent wind conditions.}

{Based on these comparisons, it is concluded that the proposed mixed FD architecture has the best performance in detecting the considered faults in all load cases, despite the fact it needs further improvement to detect $f_8$ in the wind case below the rated value (LC7) and to reduce detection delay in some fault scenarios.}

%% file: sections/5_final.tex
{FOWTs operate in the hostile marine environment with restricted accessibility and maintainability.
Currently, there is no all-encompassing FD architecture deployed on FOWTs. Such an absence poses a big challenge to reliability engineers and may potentially lead to increased O\&M costs.}
In this paper, a mixed FD architecture is established by integrating a model-based and a signal-based scheme to detect and isolate a mix of critical faults for FOWTs. 
{In order to verify the developed mixed FD architecture, a 10MW FOWT benchmark, including specific predefined faults, is developed by extending the widely-used FAST code. 
In particular, the structural faults of the rotor blades and of the mooring lines, for the first time, are taken into account for the FD purpose.}
While the model-based scheme is used to detect and isolate the actuator and sensor faults, the signal-based scheme is mainly used to detect the faults associated with the mooring lines, using only existing available measurements.

In order to illustrate the advantages of the proposed mixed FD architecture, {comparisons are} drawn with the two other classic signal-based FD approaches: PCA and DPCA. When comparing the detection and isolation of faults, results show that the proposed mixed FD architecture is able to detect and isolate critical FOWT faults in different load cases effectively. 
Compared to two classical PCA and DPCA approaches, the proposed mixed FD architecture has the best performance in fault detection of FOWTs in realistic wind and wave conditions.

{ Even though this approach shows promising results, some limitations are still lingering.
For instance, it does not take into account the isolation of the unstructured fault. In addition, the fault detection robustness, which is affected by the wind turbulence, should be further investigated.}

Based on {these discussions}, it is suggested that in the future the architecture could be extended by including a reduced-order nonlinear physical model of the 10MW FOWT benchmark, as well as a more general fault model and the capability to do fault isolation for unstructured faults. 
In addition, {the proposed mixed FD architecture can be tailored to other types of FOWTs, such as Spar type, tension leg platform type of FOWTs.
Furthermore, it can be also extended to bottom-fixed wind turbines.
Moreover}, fault tolerant control design for the 10MW FOWT benchmark should be investigated by combining the proposed mixed FD architecture with fault accommodation techniques.